\documentclass[letterpaper, 10 pt, conference]{IEEEtran}  
\pdfoutput=1

\makeatletter
\let\NAT@parse\undefined
\makeatother

\usepackage[dvipsnames]{xcolor}

\newcommand*\linkcolours{ForestGreen}

\usepackage{times}
\usepackage{graphicx}
\usepackage{amssymb}
\usepackage{gensymb}
\usepackage{amsmath}

\usepackage{subcaption}

\usepackage{url,hyperref}
\hypersetup{
colorlinks,
linkcolor=\linkcolours,
citecolor=\linkcolours,
filecolor=\linkcolours,
urlcolor=\linkcolours}
\usepackage{breakurl}
\usepackage[labelfont={bf},font=small]{caption}
\usepackage[none]{hyphenat}

\usepackage{mathtools, cuted}

\usepackage[noadjust, nobreak]{cite}

\usepackage{tabularx}
\usepackage{amsmath}

\usepackage{float}

\usepackage{pifont}

\newcolumntype{Y}{>{\centering\arraybackslash}X}

\usepackage[]{placeins}

\usepackage{placeins}

\usepackage{tikz}

\usepackage[framemethod=tikz]{mdframed}

\usepackage{afterpage}

\usepackage{stfloats}
\usepackage[switch]{lineno}
\usepackage{atbegshi}
\newcommand{\handlethispage}{}
\newcommand{\discardpagesfromhere}{\let\handlethispage\AtBeginShipoutDiscard}
\newcommand{\keeppagesfromhere}{\let\handlethispage\relax}
\AtBeginShipout{\handlethispage}

\title{\LARGE \bf
Quantum simulation of an exotic quantum critical point in a \\two-site charge Kondo circuit
}

\author
{Winston Pouse$^{1,2,3\dagger}$, Lucas Peeters$^{3,4\dagger}$, Connie L. Hsueh$^{1,2,3}$, Ulf Gennser$^{5}$, Antonella Cavanna$^{5}$, \\ Marc A. Kastner$^{4,6}$, Andrew K. Mitchell$^{7,8\ast}$, and David Goldhaber-Gordon$^{2,4\ast}$\\
\\
\normalsize{$^{1}$Department of Applied Physics, Stanford University, Stanford, CA 94305, USA}\\
\normalsize{$^{2}$Stanford Institute for Materials and Energy Sciences, SLAC National Accelerator Laboratory, Menlo Park, California 94025, USA}\\
\normalsize{$^{3}$Geballe Laboratory for Advanced Materials, Stanford University, Stanford, California 94305, USA}\\
\normalsize{$^{4}$Department of Physics, Stanford University, Stanford, CA 94305, USA}\\
\normalsize{$^{5}$ Centre de Nanosciences et de Nanotechnologies (C2N), CNRS, Univ. Paris-Sud, Université Paris-Saclay, 91120 Palaiseau, France}\\
\normalsize{$^{6}$Department of Physics, Massachusetts Institute of Technology, Cambridge, MA 02139, USA}\\
\normalsize{$^{7}$ School of Physics, University College Dublin, Belfield, Dublin 4, Ireland}\\
\normalsize{$^{8}$ Centre for Quantum Engineering, Science, and Technology, University College Dublin, Belfield, Dublin 4, Ireland}\\
\\
\normalsize{$\dagger$ These authors contributed equally to this work.}\\
\normalsize{$^\ast$To whom correspondence should be addressed;}\\ 
\normalsize{E-mail:  andrew.mitchell@ucd.ie, goldhaber-gordon@stanford.edu.}
}

\begin{document}

\maketitle
\thispagestyle{empty}
\pagestyle{empty}
\clearpage

\begin{abstract}
Tuning a material to the cusp between two distinct ground states can produce exotic physical properties, unlike those in either of the neighboring phases. The prospect of designing a model experimental system to capture such behavior is tantalizing. An array of tunnel-coupled quantum dots, each hosting a local spin, should have an appropriately complex phase diagram, but scaling up from individual dots to uniform clusters or lattices has proven difficult: though each site can be tuned to the same occupancy, each has a different set of localized wavefunctions whose couplings to neighboring sites cannot be made fully uniform. An array of metal nanostructures has complementary strengths and weaknesses: simple electrostatic tuning can make each element behave essentially identically, but intersite coupling is not tunable. In this work, we study a tunable nanoelectronic circuit comprising two coupled hybrid metallic-semiconductor islands, combining the strengths of the two types of materials, and demonstrating the potential for scalability. With two charge states of an island acting as an effective spin-½, the new architecture also offers a rich range of coupling interactions, and we exploit this to demonstrate a novel quantum critical point. Experimental results in the vicinity of the critical point match striking theoretical predictions. 
\end{abstract}

\section*{Introduction}\label{sec:intro}
The rich behaviors of bulk materials emerge from microscopic quantum interactions among their many electrons and atoms. When competing interactions favor different collective quantum states, one can often tune from one quantum state to another by applying pressure, electromagnetic fields, or chemical doping. In principle this can even happen at absolute zero temperature: a \textit{quantum phase transition}~\cite{sachdev_2011, Paschen2021}. Remarkably, the zero-temperature quantum critical point (QCP) at a specific value of a tuning parameter controls behavior over a widening range of parameter values as the temperature is increased, making signatures of criticality experimentally accessible. Further, seemingly very different systems can behave in the same `universal' way near their respective critical points.

A full microscopic description of the range and character of different phases and their transitions is in most cases impossible given the sheer chemical complexity of real bulk materials. Fortunately, simplified models often capture the essential physics of interest, providing valuable insight into the behavior of bulk materials and even guiding design of new materials. Typically, these models describe a set of local sites, each hosting one or a few interacting quantum degrees of freedom, coupled to other sites and sometimes to conducting reservoirs~\cite{Anderson1961, Georges1995, affleck2009quantum, Gull2011}. Calculating the low-energy properties of even these simplified models on clusters of more than a few sites exceeds the capabilities of the most powerful classical computers. Digital quantum computers could work for such calculations, but only once they are scaled to a far greater number of quantum bits than the present state of the art. Highly tunable nanoelectronic circuits based on one or a few semiconductor quantum dots can act as analog quantum simulators, directly implementing Hamiltonians of interest and thus offering the near-term prospect of more powerful computation than other currently available approaches. These circuits display diverse phenomena including Coulomb blockade~\cite{Meirav1990}, various Kondo effects~\cite{Goldhaber-Gordon1998_Nature,  Cronenwett1998, Sasaki2000, Jeong2001, Oreg2003, Potok2007, Buizert2007, Takada2014}, emergent symmetries~\cite{Keller2013,mitchell2021so}, and fractionalization~\cite{Keller2015,Iftikhar2015,Iftikhar2018,landau2018charge}. Quantum phase transitions with universal properties have also been realized in such circuits~\cite{Keller2015,Iftikhar2015, mitchell2016universality, Iftikhar2018, zhang2021}.

However, the circuits studied so far cannot fully capture collective behavior over many sites of a lattice~\cite{burdin2000coherence, Stafford1994}. A long-standing goal is to scale up these circuits to more directly mirror the structure of bulk materials. For example, a four-site Fermi-Hubbard system was recently studied using semi-automated control and tuning capabilities~\cite{Hensgens2017}, but scaling up to a larger uniform lattice is daunting. Even with advanced modern fabrication, disorder in doping and lithographic patterning can make the level spectra of two nominally identical quantum dots inequivalent~\cite{Pateras2019}. An array of gates may be used to equalize local chemical potentials across different sites, but many-body ground states are still affected by the full level spectrum of each site~\cite{Inoshita1993}, which cannot be fully controlled in these systems.

 A recently-introduced paradigm for quantum simulation of quantum phase transitions is based on a local site formed from a hybrid metal-semiconductor island. Here we demonstrate  that this approach is uniquely advantageous for scaling to larger arrays: because the metal component hosts an effective continuum of single particle states, different islands of the same size behave essentially identically. As a step toward such scaling, the nature of the coupling between neighboring islands must be understood. In this article, we develop a model to describe the inter-island coupling in the simplest two-site system, and validate this model experimentally using transport measurements. We show that this device architecture generates a new kind of correlated tunneling interaction, which can drive a collective (many-body coherent) screening of the two sites. Furthermore, this interaction can compete with the usual island-lead Kondo effects, resulting in a novel quantum critical point. Measured conductance signatures across the quantum phase transition – which is a variant of the long-sought two-impurity Kondo transition – match universal theoretical predictions for the model. Scaling up to many such coupled sites will then allow experimental simulation of lattice models that are beyond the reach of traditional computational techniques.

\begin{figure*}[t]
\centering
\begin{subfigure}{\linewidth}
  \centering
  \includegraphics[width=5 in]{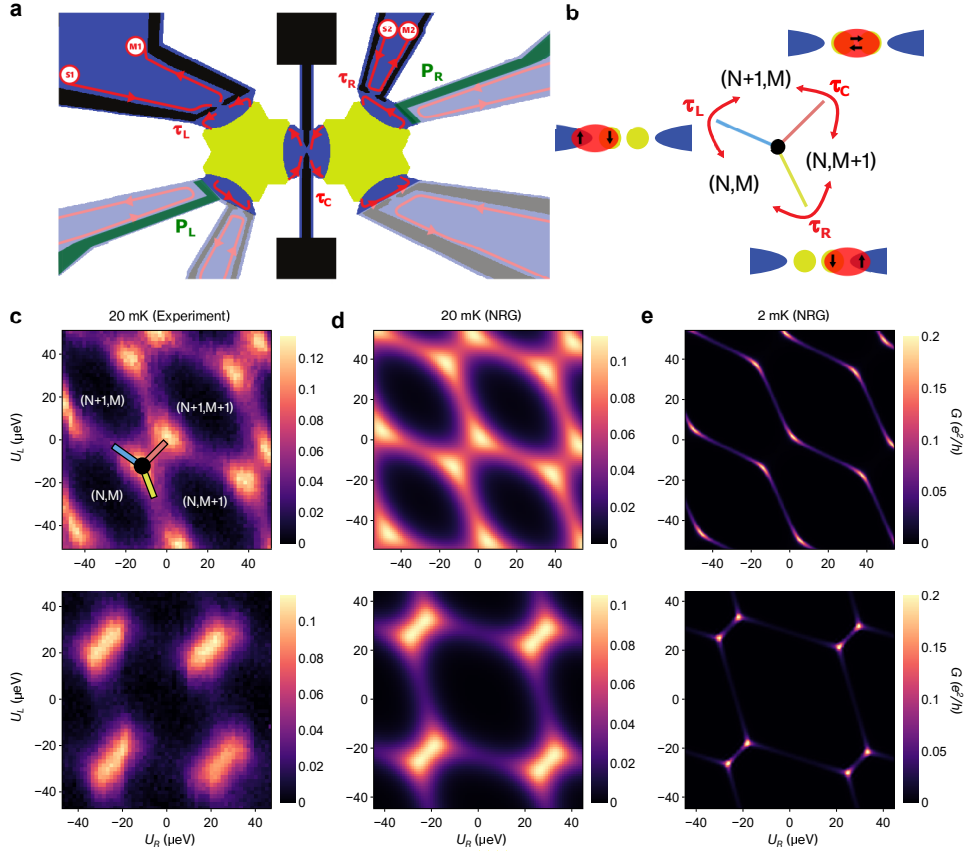}
  \phantomcaption
  \label{fig:schematic}
\end{subfigure}
\begin{subfigure}{0\textwidth}
  \centering
  \includegraphics[width=\linewidth]{RoughDraftFigures/fig1Small.eps}
  \phantomcaption
  \label{fig:cartoon}
\end{subfigure}
\begin{subfigure}{0\linewidth}
  \centering
  \includegraphics[width=.98176\linewidth]{RoughDraftFigures/fig1Small.eps}
  \phantomcaption
  \label{fig:stabExpt}
\end{subfigure}
\begin{subfigure}{0\linewidth}
  \centering
  \includegraphics[width=.95\linewidth]{RoughDraftFigures/fig1Small.eps}
  \phantomcaption
  \label{fig:stabNRG20}
\end{subfigure}
\begin{subfigure}{0\linewidth}
  \centering
  \includegraphics[width=.9739\linewidth]{RoughDraftFigures/fig1Small.eps}
  \phantomcaption
  \label{fig:stabNRG02}
\end{subfigure}\vspace{-15pt}
\caption{\textbf{Two island charge-Kondo device.} \textbf{a,} Schematic layout of the device structure, consisting of two metallic islands (yellow-green) coupled to quantum Hall edges (red lines) in a buried 2DEG (blue) via QPCs (black). Only the top and central QPCs are used throughout this work. The island levels are controlled via plunger gates (green).  \textbf{b,} Three neighbouring charge states are interconverted by direct tunneling of electrons at each of the three QPCs, characterized by transmissions $\tau_L, \tau_R, \tau_C$. Distinct Kondo effects arise along each two-state degeneracy line. At the triple point connecting them, the three different Kondo interactions cannot simultaneously be satisfied, leading to frustration and a quantum critical point. \textbf{c, d,} Experimentally measured series conductance and NRG calculations at 20 mK for $\tau_L=\tau_R \equiv \tau=0.38$ ($J_L = J_R \equiv J=0.35$) as the island potentials $U_L, U_R$ are varied via plunger gate voltages $P_L$ and $P_R$. The top row corresponds to $\tau_C = 0.9$ ($J_C = 0.5$) and the bottom row $\tau_C = 0.7$ ($J_C=0.3$). The bright conductance spots in the top row correspond to the triple points. In the bottom row, the triple points are closer and somewhat merged. \textbf{e,} NRG calculated stability diagram at 2 mK for the same settings as in \textbf{d}. While experimentally inaccessible, we see clear peaks at the triple points, with suppressed conductance elsewhere.}
\label{fig:device}
\end{figure*}

Our device consists of a circuit containing two coupled hybrid metal-semiconductor islands, each also coupled to its own lead, as illustrated schematically in Fig.~\ref{fig:schematic}. Even though the islands are small enough to have a  substantial charging energy, the metal component endows each island with an effective continuum of single particle states. This contrasts with the discrete and individualized level spectrum of purely semiconductor quantum dots noted above. Our circuit is based on a GaAs/AlGaAs heterostructure which hosts a buried two-dimensional electron gas (2DEG). Mesas are lithographically patterned (blue regions in Fig.~\ref{fig:schematic}, outside of which the 2DEG is etched away). Metallic islands are deposited bridging the various mesas, then are electrically connected to the 2DEG by thermal annealing. The device is operated in a magnetic field of $4.3$~T, corresponding to a quantum Hall filling factor $\nu=2$ in the 2DEG bulk. The left and right islands are designed to behave identically: the spacing of single-particle states on each island is far below $kT$ at our base temperature of $20$ mK; and their charging energies $E_C^L\approx E_C^R\approx 25~\mu$eV are equal to within our experimental resolution ($V\approx 10~\mu$eV is the inter-island capacitive interaction). Lithographically patterned metallic top gates form quantum point contacts (QPCs, black in Fig.~\ref{fig:schematic}). The transmissions $\tau_L$ and $\tau_R$ control the left and right island-lead tunnel couplings, while $\tau_C$ controls the coupling between the islands. Each coupling is through the outermost quantum Hall edge state; QPC voltages are set so that the second, inner channel, is completely reflected. Throughout the experiment we fix $\tau_L=\tau_R\equiv \tau$ and keep all other QPCs closed. Finally, plunger gates (green) control the electrostatic potential, and hence electron occupancy, on each island. We measure the conductance $G$ from left lead to right lead through both islands in series, as a function of the left and right plunger gate voltages $P_L$ and $P_R$. See Methods for further details of the device and measurement setup.

Following the charge-Kondo mapping for a single island introduced theoretically by Matveev~\cite{Matveev1995}, and validated experimentally by Iftikhar~\cite{Iftikhar2015,Iftikhar2018}, we formulate an effective model describing our two-site device. Our double charge-Kondo (DCK) model reads,
\begin{equation}\label{eq:H2dck}
\begin{split}
&H_{\rm DCK} = H_{\rm elec} + \left ( J_L^{\phantom{\dagger}} \hat{\mathcal{S}}^+_L \hat{s}^-_L + J_R^{\phantom{\dagger}} \hat{\mathcal{S}}^+_R \hat{s}^-_R  + {\rm H.c.} \right ) \\
 &+ \left (J_C^{\phantom{\dagger}} \hat{\mathcal{S}}^+_L \hat{\mathcal{S}}^-_R c_{C_L}^{\dagger}c_{C_R}^{\phantom{\dagger}} +{\rm H.c.} \right ) + I \hat{\mathcal{S}}^z_L \hat{\mathcal{S}}^z_R + B_L^{\phantom{z}} \hat{\mathcal{S}}^z_L + B_R^{\phantom{z}}\hat{\mathcal{S}}^z_R\;.
\end{split}
\end{equation}
where $H_{\rm elec}$ describes the independent conduction electron reservoirs around each of the three QPCs, $\hat{\mathcal{S}}^+_{L(R)}$, $\hat{\mathcal{S}}^-_{L(R)}$, $\hat{\mathcal{S}}^z_{L(R)}$ are pseudospin$\tfrac{1}{2}$ operators for the left (right) site, $\hat{s}^+_{L(R)}$ and $\hat{s}^-_{L(R)}$ are pseudospin-$\tfrac{1}{2}$ raising and lowering operators for the corresponding electronic reservoirs, and $c_{C_L}^{\dagger}c_{C_R}^{\phantom{\dagger}}$ represents electronic tunneling from the right to the left site. The terms in the first set of parentheses favour Kondo screening of the pseudospin on each site by its attached lead, whereas those in the second set of parentheses represent pseudospin coupling between the two sites. Finally, the term proportional to $I$ describes an inter-island capacitive interaction, while individual ``plunger’’ gates capacitively coupled to the two sites create local Zeeman pseudospin fields $B_L$ and $B_R$. Larger excursions of plunger gate voltages add or subtract electrons from the two sites, which is not captured in this effective spin-$\tfrac{1}{2}$ Hamiltonian.

This effective Hamiltonian associates effective pseudospin degrees of freedom to the macroscopic island charge states. The raising/lowering operator $\hat{\mathcal{S}}_{L}^{\pm}$ converts a state $(N,M)$ with $N$ electrons on the left island and $M$ on the right, into the state $(N\pm 1,M)$; while $\hat{\mathcal{S}}_{R}^{\pm}$ yields $(N,M\pm 1)$. These transitions naturally arise from single-electron tunneling through the left and right QPCs, respectively. Similarly, we see that the process that interconverts $(N,M+1)$ and $(N+1,M)$, the novel ingredient of the DCK model, is precisely the tunneling of an electron through the central QPC. This provides the $c_{C_L}^{\dagger}c_{C_R}^{\phantom{\dagger}}$ term, but because each island's individual charge occupation is mapped onto a pseudospin, the tunneling also causes the mutual pseudospin flip, $\hat{\mathcal{S}}^+_L \hat{\mathcal{S}}^-_R$. 
These processes are illustrated in Fig.~1b.

The DCK model is reminiscent of the two-impurity Kondo (2IK) model, which captures the competition between Kondo screening of local moments and RKKY exchange interaction. However, the DCK model has a major difference: the inter-site coupling is not a simple exchange interaction but rather a correlated tunneling. The concerted pseudospin flip of the sites is accompanied by electronic tunneling at the central QPC. This favours the formation of an inter-site Kondo singlet with many-body entanglement, rather than the simple two-body local spin singlet arising from an RKKY interaction. When scaled to a lattice of sites, this new interaction may produce the lattice coherence effect seen in heavy fermion materials but not so far accounted for in microscopic models~\cite{burdin2000coherence}. 

A crucial feature of the present charge-Kondo implementation is that the pseudospin couplings $J_L$, $J_R$ and $J_C$ in the DCK model Eq.~\ref{eq:H2dck} are related directly to the QPC transmissions of the device $\tau_L$, $\tau_R$ and $\tau_C$, and can be large. By tuning these couplings, one can realize various Kondo effects, and indeed a quantum critical point, at relatively high temperatures. This contrasts with the more familiar coupling of spins between two semiconductor quantum dots, where the effective exchange interactions are perturbatively small, being derived from an underlying Anderson model. Furthermore, relevant perturbations present in the Anderson model destroy the quantum phase transition of the oversimplified 2IK model, replacing it with a smooth crossover~\cite{Zarand2006, jayatilaka2011two}. The two island charge-Kondo system therefore presents a unique opportunity to observe a two-impurity QCP at experimentally relevant temperatures.


\section*{Results}
\subsection*{Phase diagram and Kondo competition}\label{sec:pd}
The island charging energies $E_C^{L,R}$ and inter-island capacitive interaction $V$ are finite in the physical device, so multiple island charge states play a role. This gives rise to a periodic hexagonal structure of the charge stability diagram as a function of the left and right plunger gate voltages $P_{L,R}$. We convert these to energies $U_{L,R}=U_{L,R}^0+\alpha P_{L,R}$ using the experimentally measured capacitive lever arm ${\alpha=50}~\mu\text{eV/mV}$, relative to an arbitrary reference $U_{L,R}^0$. $U_{L,R}$ are related to the pseudo-Zeeman fields $B_{L,R}$ of $H_{\rm tune}$ in Eq.~\ref{eq:H2dck} via $\vec{B}=\bar{\boldsymbol{\alpha}} \vec{U}$, which  accounts for cross-capacitive gate effects. 

The experimental stability diagram in Fig.~\ref{fig:stabExpt} allows us to identify regimes with particular charge states on the two islands. In particular, we see distinct charge degeneracy lines $(N,M)/(N,M+1)$, $(N,M)/(N+1,M)$ and $(N+1,M)/(N,M+1)$, each of which is associated with single electron tunneling at one of the three QPCs (see Fig.~\ref{fig:cartoon}). This structure, including its characteristic gate periodicity, is reproduced very well by numerical renormalization group (NRG) \cite{bulla2008numerical,mitchell2014generalized} calculations of the DCK model,  generalized to take into account multiple charge-state on each island (see Methods and Supplementary Info). We fit $J_{L,R,C}$ for a given set of experimental transmissions $\tau_{L,R,C}$, as shown in Fig.~\ref{fig:stabNRG20} for the same temperature ${T=20}$~mK. The stability diagrams are periodic over an even larger range of gate voltages than shown, but in general we limit the range to minimize drift and to make efficient use of measurement time as we explore a many-dimensional parameter space. 

Along the degeneracy line $(N,M)/(N+1,M)$ the left island charge pseudospin is freely flipped by tunneling at the QPC between the left island and lead, giving rise to a Kondo effect due to the first term of Eq.~\ref{eq:H2dck}. However, the series conductance from left to right leads through the double island structure is suppressed by this effect, since the conductive pathway involves virtual polarization of the Kondo singlet through the state $(N,M+1)$, which is an excited state when the charge dynamics of the right island are frozen. This is supported by NRG calculations at ${T=2}$~mK (Fig.~\ref{fig:stabNRG02}) which show this incipient `Kondo blockade'~\cite{mitchell2017kondo}  in the series conductance. A similar effect is seen along the degeneracy line $(N,M)/(N,M+1)$, which corresponds to a Kondo effect of the right island with the right lead. Along the degeneracy line $(N+1,M)/(N,M+1)$, tunneling with the leads is not involved. Instead we may regard $(N+1,M)$ and $(N,M+1)$ as two components $\Leftarrow$ and $\Rightarrow$ of a collective pseudospin state of the double island structure, which is flipped by electronic tunneling at the central QPC. This gives rise to a new kind of inter-island Kondo effect. The resulting Kondo singlet is disrupted by tunneling at the leads, and hence the low temperature conductance is again suppressed. At first glance, the 2D plots of conductance as a function of gate-tuned occupancy of the two sites look reminiscent of those measured on conventional semiconductor double quantum dot systems. Below we mention some important differences from this more familiar situation. For a somewhat more extensive discussion on how the low temperature behavior is distinct from that of a double quantum dot, see Supplementary Info. 


\subsection*{Triple point}
The triple point (TP), where $(N,M)/(N,M+1)/(N+1,M)$ are all degenerate, is a special point in the phase diagram. Here the three Kondo effects described above are all competing, see Fig.~\ref{fig:cartoon}. At $J_L=J_R=J_C$ in Eq.~\ref{eq:H2dck}, the resulting frustration gives rise to a QCP, which will be the main focus of this work.  

At the TP, the high-temperature series conductance is enhanced because an electron can tunnel from left lead to right lead through the islands without leaving the ground state charge configurations. Neglecting interactions and treating the QPCs as three resistors in series, the maximum conductance $G=e^2/3h \equiv G^*$ occurs when the tunneling rates at each constriction are equal, $J_L=J_R=J_C$, since then there is no bottleneck in the flow of electrons through the structure. When the three QPCs are opened up, making the energy scales for tunneling large compared to our base temperature, $G\simeq e^2/3h$ is indeed observed experimentally. This compares to a maximum series conductance $2e^2/h$ or $e^2/h$ in a conventional semiconductor double quantum dot (with or without spin). Importantly, we also find that in our full model including interactions $G=G^*$ at a TP at low temperature $T/T_K\ll 1$, provided $J_L=J_R=J_C \equiv J$.  Here, $T_K\sim E_C \exp[-1/2\nu J]$ is the Kondo temperature at the QCP (with $\nu$ the electronic density of states at the Fermi energy). In this case, we find from NRG that the conductance increases with decreasing $T$, approaching the critical value as $G^* - G(T) \sim (T/T_K)^{2/3}$. Based on this and the unusual residual $T=0$ entropy $\Delta S=\ln(\sqrt{3})$ seen in NRG calculations, we conclude that the QCP hosts a non-Fermi liquid with exotic \textit{fractional} excitations. In contrast, as noted above, no QCP exists in a conventional double quantum dot; unavoidable charge fluctuations between the dots blur the phase transition studied in the 2IK model into a broad crossover~\cite{Zarand2006, jayatilaka2011two}.


\begin{figure}[ht!]
    \centering
    \begin{subfigure}{\columnwidth}
        \centering
        \includegraphics[width=3.4 in]{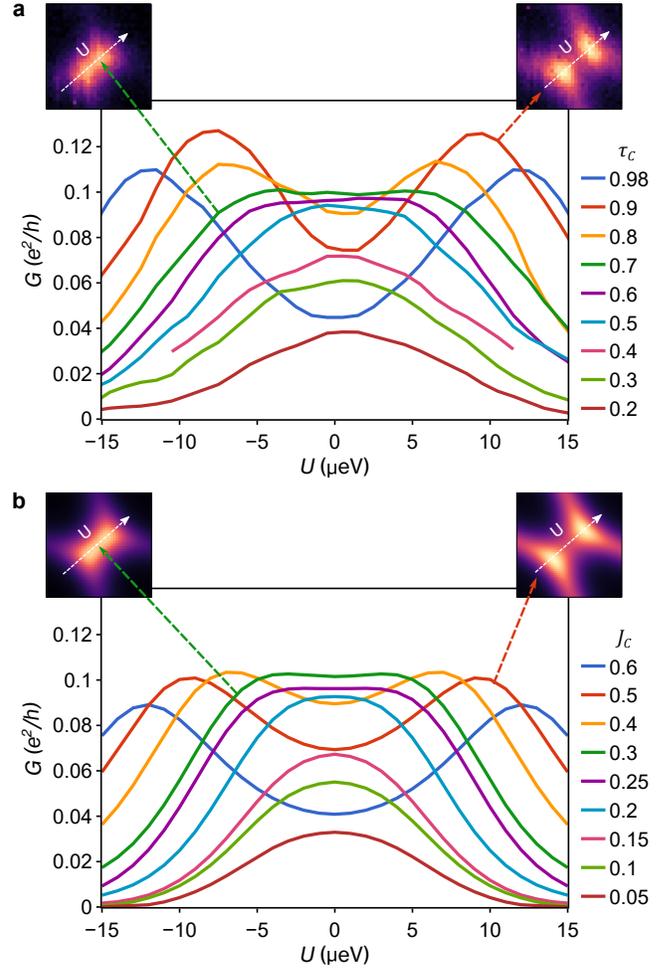}
        \phantomcaption
        \label{fig:linecut20}
    \end{subfigure}\ignorespaces
    \begin{subfigure}{0\columnwidth}
        \centering
        \includegraphics[width=\columnwidth]{RoughDraftFigures/fig2Unsymm.eps}
        \phantomcaption
        \label{fig:linecutcalc}
    \end{subfigure}\ignorespaces
\caption{\textbf{Conductance line cuts between triple points.} Experimentally-measured (\textbf{a}) and NRG-calculated (\textbf{b}) line cuts for $\tau=0.38$ ($J=0.35$ in the model) along the line $U_L=U_R\equiv U$ for different $\tau_C$ ($J_C$). Insets show representative 2D $P_L,P_R$ sweeps from which line cuts are extracted. The model parameters in \textbf{b} are optimized to fit the experiment (see Supplementary Info.).}
\label{fig:linecut20comp}
\end{figure}

\subsection*{Conductance Line Cuts}
We first focus on the behavior of conductance near the TPs. Specifically, in Fig.~\ref{fig:linecut20} we take cuts along the line between TPs, $U_L=U_R\equiv U$, for different $\tau_C$ at fixed $\tau_L=\tau_R\equiv \tau$. $U=0$ is chosen to be the high-symmetry point between TPs. Experimental data are compared with the corresponding NRG simulations of the device in Fig.~\ref{fig:linecutcalc}. Since the physics of DCK only relies on charge states, every pair of triple points is indistinguishable, and we extract the experimental line cuts of Fig.~\ref{fig:linecut20} from averages over multiple pairs of triple points. 

Experiment and theory are seen to match very well, notably with regard to the conductance values at all $U$ for most $\tau_C$ ($J_C$), the width of the peak for $\tau_C \leq 0.7$ ($J_C \leq 0.4$), and the positions of the split peaks at the largest $\tau_C$ ($J_C$). This level of agreement validates the use of the DCK model to describe the physical device.  

We note that the TP positions in the space of $(U_L,U_R)$ depend on $\tau_C$. At large $\tau_C$, the peaks are rather well separated and can be easily distinguished, but due to temperature broadening the peaks are found to merge at low $\tau_C$. Within NRG, this effect is reproduced on decreasing $J_C$. 

Although the TP positions at low $J_C$ can still be identified in NRG by going to lower temperatures where the peaks sharpen up (compare Figs.~\ref{fig:stabNRG20} and \ref{fig:stabNRG02}), even at the experimental base electron temperature of 20 mK, thermal broadening complicates the experimental analysis of the TP behavior. Thus care must be taken to estimate the TP positions from the full stability diagram and to disentangle the influence of adjacent TPs.

We also see clear non-monotonicity in the experimental conductance as a function of $\tau_C$, reflecting the competition between the different Kondo interactions at the TP. Taking the critical point with completely frustrated interactions to be at $\tau_C^*$ (a monotonic function of $\tau$, and not necessarily $\tau_C^*=\tau$), we expect lower conductance when the ground state is a Fermi-liquid Kondo singlet -- either where the island-lead Kondo effects dominates, $\tau_C<\tau_C^*$, or the inter-island Kondo effect dominates, $\tau_C>\tau_C^*$ (recall Fig.~\ref{fig:cartoon}). This is because the dominant Kondo interaction suppresses conductance through the other channel(s) in series, blocking series transport across the whole device. This does not occur in a conventional DQD, where a physically simpler singlet (analogous to RKKY, not Kondo) links the spins on the two dots. This coupling has no strong temperature dependence, unlike what we observe experimentally and capture in NRG.

In our experiments, conductance is observed to be higher for $\tau_C\sim \tau_C^*$ as the frustration of competing Kondo interactions enhances charge fluctuations that would normally be frozen out at low temperatures. This feature is not well-captured by NRG calculations at these temperatures, nor would it be expected to be. At large $\tau_C$, many charge states on the islands contribute to transport, but within NRG the number of states that can be included in the calculations in practice is limited (Supplementary Info.), even when using a supercomputer. 

Despite some uncertainty in the precise TP location, the very existence of a QCP implies an underlying universality, in terms of which conductance signatures in its vicinity can be quantitatively analyzed.

\begin{figure*}[ht!]
    \centering
    \begin{subfigure}[h]{\linewidth} 
        \includegraphics[width=7 in]{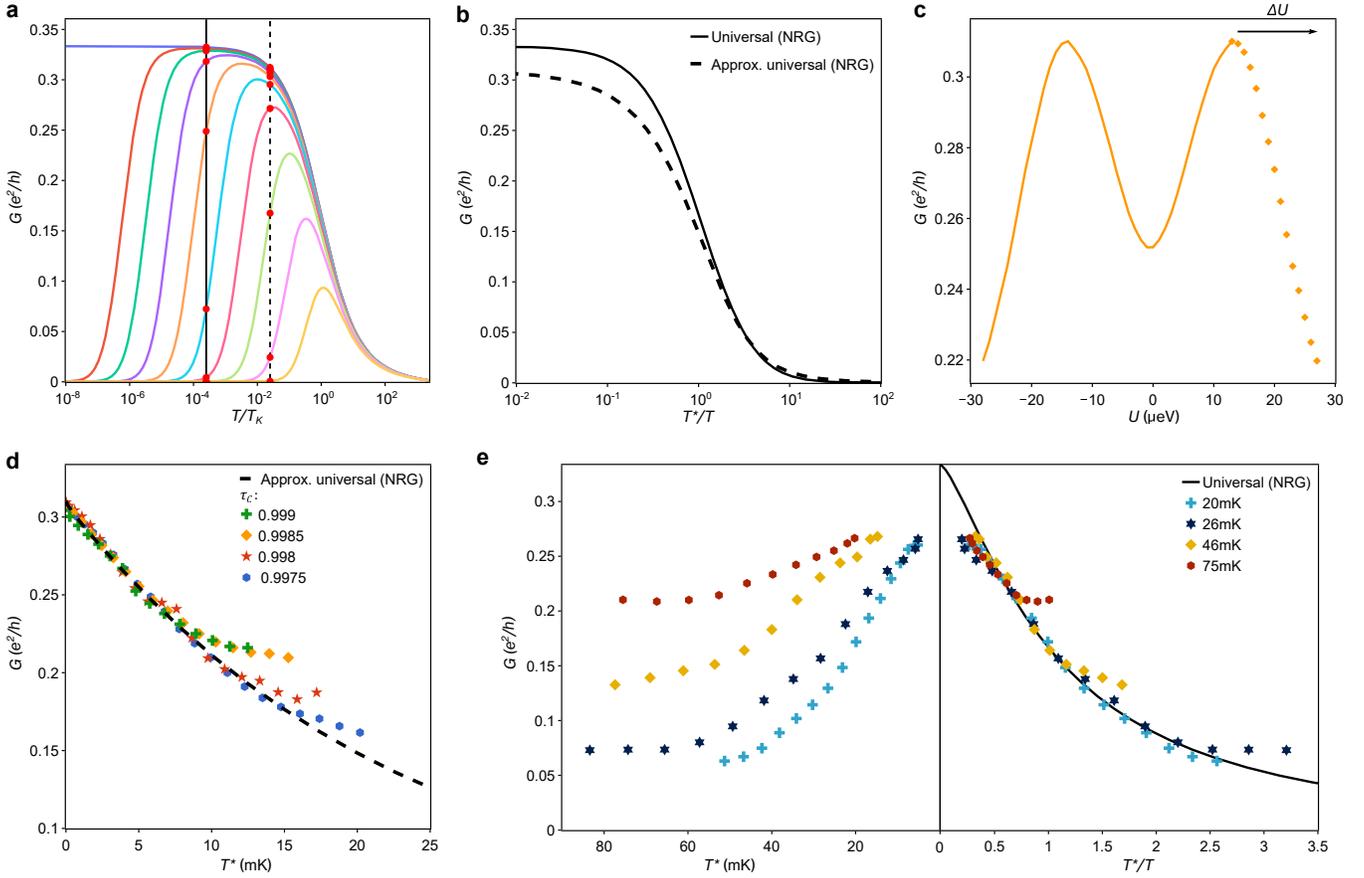}
        \phantomcaption
        \label{fig:energyScales}
    \end{subfigure}
    \begin{subfigure}[t]{0\textwidth} 
         \includegraphics[width=\textwidth]{RoughDraftFigures/fig4Updated.eps}
         \phantomcaption
         \label{fig:universalCurve}   
    \end{subfigure}
    \begin{subfigure}[t]{0\textwidth} 
         \includegraphics[width=\textwidth]{RoughDraftFigures/fig4Updated.eps}
         \phantomcaption
         \label{fig:trplDetuning}   
    \end{subfigure}
    \begin{subfigure}[t]{0\textwidth} 
         \includegraphics[width=\textwidth]{RoughDraftFigures/fig4Updated.eps}
         \phantomcaption
         \label{fig:tauCollapse}   
    \end{subfigure}
    \begin{subfigure}[t]{0\textwidth} 
         \includegraphics[width=\textwidth]{RoughDraftFigures/fig4Updated.eps}
         \phantomcaption
         \label{fig:tempCollapse}   
    \end{subfigure}\vspace{-10pt}
\caption{\textbf{Universal conductance scaling near triple point.} \textbf{a.} Theoretically calculated conductance as a function of $T/T_K$ for different $T^*$ (increasing from the left, blue colored curve, to the right, yellow curve) using a minimal model valid near the TP. The intersection with the solid (at $T/T_K = 2.5 \times 10^{-4}$) and dashed (at $T/T_K = 2.5 \times 10^{-2}$) lines represent ways to reconstruct the universal and approximately universal curves of \textbf{b}. \textbf{b.} Theoretical universal curve (solid line) as a function of $T^*/T$ in the limit $T/T_K \ll 1$, compared with an approximately universal curve obtained for $T/T_K\lesssim 1$ (dashed line). \textbf{c.} Measured line cut between triple points similar to that in Fig.~\ref{fig:linecut20comp}, except here $\tau=0.95$ and $\tau_C\approx 0.9985$. $T^*$ is a function of gate-induced detuning $\Delta U$ away from a TP (Eq.~\ref{eq:Tstar}), which we define relative to $U_{\rm TP}$ estimated at the peak. \textbf{d.} We plot the truncated line cut of \textbf{c} as well as similar line cuts (symbols) for different $\tau_C$ as a function of $T^*$. These fall on top of each other and the approximately universal curve of \textbf{b}. The uptick in conductance at the tail of each cut results from influence of a neighboring TP not included in the minimal model. \textbf{e.} Measured truncated line cuts (symbols) at different temperatures are plotted in the left panel. From lowest to highest temperatures, $\tau = \left\{0.78, 0.78, 0.81, 0.82\right\}$ and $\tau_C = 0.9$ (Methods). Due to a detuning $\Delta \tau_C$, Eq.~\ref{eq:Tstar} implies $T^*>0$ even at $\Delta U=0$. This shift is estimated in fitting the tails to the universal curve in the right side when plotted as a function of $T^*/T$, whence we observe universal scaling collapse.}
\label{fig:scalingCollapse}
\end{figure*}

\subsection*{Universal Scaling}
We now turn to the behavior near the QCP, resulting from frustrated island-lead and inter-island Kondo effects. We focus on parameter regimes with large $\tau$ and $\tau_C$, such that the corresponding Kondo temperatures are large. This allows experimental access to the universal regime $T/T_K \lesssim 1$. At the QCP with $\tau_C \simeq \tau_C^*$, our theoretical analysis predicts ${G\simeq G^*=e^2/3h}$. Moreover, non-trivial behavior is observed in the vicinity of this singular point, where perturbations drive the system away from the QCP and towards a regular Fermi liquid state. The associated conductance signatures are entirely characteristic of the quantum phase transition in this system.

Since the low-$T$ physics near a QCP is universal and therefore insensitive to microscopic details, we use a minimal model, with only $(N,M)/(N+1,M)/(N,M+1)$ states retained in Eq.~\ref{eq:H2dck}. The QCP is destabilized by either detuning the couplings, corresponding to the perturbation $\Delta \tau_C = \tau_C-\tau_C^*$, or moving away from the TP via $\Delta U = U-U_{\rm TP}$ (where $U_{\rm TP}$ is the putative TP position). Remarkably, we find from NRG that any combination of $\Delta \tau_C$ and $\Delta U$ can be captured by a \emph{single} scale $T^*$ characterizing the flow away from the QCP, provided the magnitude of the perturbations is small:
\begin{equation}\label{eq:Tstar}
T^* = a|\Delta \tau_C|^{3/2} + b|\Delta U|^{3/2}
\end{equation}
with $a,b$ constants. The resulting universal conductance curve as a function of $T^*/T$ is shown as the solid line in Fig.~\ref{fig:universalCurve}. In practice, this can be reconstructed by fixing a small $T/T_K \ll 1$ and varying $\Delta U$, say, while keeping $\Delta \tau_C=0$, as shown in Fig.~\ref{fig:energyScales} (left vertical solid black line, $T/T_K = 2.5 \times 10^{-4}$). Repeating the procedure at a higher temperature (right vertical dashed black line, $T/T_K = 2.5 \times 10^{-2}$), where $T/T_K \ll 1$ is not well satisfied, leads to the approximately universal curve shown as the dashed line in Fig.~\ref{fig:universalCurve}. This curve depends both on $T^*/T$ as well as $T/T_K$; the latter deduced from experimental data from the saturation value of the conductance.

We use these NRG results to interpret the experimental data. To do this we must identify the TP position, and hence $\Delta U$ from the line cut data. This is relatively straightforward when the TPs are well separated, as happens at large $\tau_C$, and we simply take $U_{\rm TP}$ as the peak position, Fig.~\ref{fig:trplDetuning}.

Since the limit $T/T_K \ll 1$ is not perfectly satisfied in experiment, we use the approximately universal curve to demonstrate the scaling collapse of data for different $\tau_C$ using Eq.~\ref{eq:Tstar} in Fig.~\ref{fig:tauCollapse} (Methods). The small $T^*$ behavior falls onto the theory curve for all $\tau_C$ considered, revealing the non-trivial $3/2$ power law scaling in the data (see Methods for further analysis). Deviations from the line are due to influence of the neighboring TP. 

Finally, we demonstrate that $T^*/T$ is indeed the universal scaling parameter by measuring and rescaling line cuts at different temperatures. In the left panel of Fig.~\ref{fig:tempCollapse}, we plot line cuts at different temperatures for slightly lower fixed values of $\tau, \tau_C$ than in Fig.~\ref{fig:tauCollapse}, as a function of $U$. At each temperature, $\tau, \tau_C$ are adjusted such that the conductance at $\Delta U = 0$ is roughly the same. We do not expect to satisfy $\tau_C = \tau_C^*$ for each temperature. Remarkably, the additive form of the contributions to $T^*$ in Eq.~\ref{eq:Tstar} means that it is unnecessary to be exactly at the critical value of $\tau_C$: the detuning in $\tau_C$ simply generates a finite $T^*$ even when $\Delta U=0$, which we can account for by a simple shift when plotting the data in terms of $T^*$ (Methods).

The right panel of Fig.~\ref{fig:tempCollapse} shows the same data scaled now as $T^*/T$, and compared with the fully universal NRG curve from Fig.~\ref{fig:universalCurve}. We cannot make the same comparison to an approximately universal curve as $T/T_K$ would change for each temperature. Instead, we understand the origin of the deviations at low $T^*/T$ and expect agreement at larger $T^*/T$. 

Indeed, the collapse and strong quantitative agreement with the non-trivial universal conductance curve is both striking and a direct signature of the novel critical point. Significantly, the collapse is over the entire range of $T^*/T$ for each line cut, with limitations at small $T^*/T$ due to the finite $T/T_K$ mentioned previously and at large $T^*/T$ due to an increasing overlap with other TPs. The observed scaling collapse for different temperatures is a result of the universal form of the conductance as a function of $T^*/T$ as well as the $3/2$ power law scaling of $T^*$ with $\Delta U$ that arises near the QCP.

\section*{Discussion}

In this work, we have presented strong evidence for a completely novel quantum phase transition in a two-site circuit. By exploiting the charge-Kondo paradigm, our device maps to a variant of the celebrated two-impurity Kondo model, here featuring a new phase in which the local moments on the two islands are screened \textit{collectively} by many-body effects driven by conduction electron scattering. This may have relevance for the emergence of lattice coherence in Kondo lattice systems.

We formulate a new model to describe the two island charge-Kondo device, and demonstrate quantitative agreement between NRG calculations and experimentally measured conductance, including in the universal regime of the exotic quantum critical point.

Our work on the crucial role of the inter-island interaction paves the way for a host of other studies. Opening each of the islands to a second lead (already present but not used in the existing device) would produce two sites each hosting a two-channel Kondo (2CK) state, and would allow studying the effect of coupling the two. Conceivably, the well-localized Majorana zero mode associated with a 2CK state at low temperatures on one island \cite{Emery1992, Coleman1995, Iftikhar2018} could be shifted to the other island by gate voltage tuning (presumably through an as-yet-unknown intermediate collective state).

Unlike for tunnel-coupled semiconductor quantum dots, there is no clear roadblock to scaling this platform to more complex uniform clusters of coupled charge-Kondo islands, and ultimately lattices. This provides a way of examining with unprecedented control some of the most subtle collective dynamics of real correlated materials, and introducing a flexible set of effective interactions.
Such scaled-up charge-Kondo clusters would act as analog quantum simulators with capabilities beyond classical computation: three coupled islands is already out of reach for NRG, while stochastic algorithms such as Quantum Monte Carlo (QMC)~\cite{Gull2011} may not be able to access the universal low-temperature dynamics of these systems. Indeed, tunable analog quantum simulators of this type may eventually form the basis for calculations requiring solutions of complex cluster models that are difficult to obtain using NRG or even QMC, as arise for example as inputs to extensions of dynamical mean field theory (DMFT)~\cite{Maier2015,senechal2015} for correlated materials such as the high-temperature superconductors.


\section*{Data Availability}
\noindent All data used in this work are available in the Stanford Digital Repository at https://doi.org/10.25740/mx151nn9365. 

\section*{Acknowledgements}
\noindent We thank F. Pierre, I. Safi, G. Zarand, C.P. Moca, I. Weymann, P. Sriram, E. Sela, Y. Oreg, Q. Si, and C. Varma for their scientific insights and suggestions. To make this project work, before coupling two islands we had to start by reproducing F. Pierre's tour de force experiments on single islands of the same type. F. Pierre helped with comments on our fabrication process, measurement procedure, and analysis. We acknowledge G. Zarand, C.P. Moca, and I. Weymann for early discussions of the Hamiltonian and its implications. Measurement and analysis were supported by the U.S. Department of Energy (DOE), Office of Science, Basic Energy Sciences (BES), under Contract DE-AC02-76SF00515. Growth and characterization of heterostructures was supported by the French Renatech network. Theory and computation (A.K.M.) were supported by the Irish Research Council Laureate Awards 2017/2018 through Grant No. IRCLA/2017/169. Part of this work was performed at the Stanford Nano Shared Facilities (SNSF), supported by the National Science Foundation under award ECCS-2026822. Early research that established how to meet the demanding technical conditions for sample fabrication and basic measurements was supported by the National Science Foundation (NSF) under award no. 1608962. W.P. acknowledges support from the Fletcher Jones Fellowship. C.L.H. acknowledges support from the Gabilan Fellowship. L.P. acknowledges support of the Albion Walter Hewlett Fellowship.

\section*{Author Contributions}
\noindent W.P. and L.P. performed the measurements. L.P. fabricated the device. A.K.M. developed the theory and did NRG calculations. W.P., L.P., C.L.H., M.A.K., A.K.M., and D.G.-G. analyzed the data. A.C. and U.G. grew the heterostructure that hosts the 2DEG on which these samples are built. D.G.-G. supervised the project.

\section*{Competing Financial Interests}
\noindent The authors declare no competing financial interests.

\newpage

\section*{Methods}\label{sec:methods}
\subsection*{Theoretical Modeling of the Device}\label{sec:methods_model}
Left and right islands are coupled to their respective leads and to each other by QPCs. We define independent electronic systems around each QPC corresponding to the blue regions in Fig.~\ref{fig:schematic},
\begin{equation}\label{eq:Helec}
H_{\rm elec} = \sum_{\alpha,\gamma,k} \epsilon_{\alpha\gamma k}^{\phantom{\dagger}} c_{\alpha\gamma k}^{\dagger}c_{\alpha\gamma k}^{\phantom{\dagger}} \;,
\end{equation}
where $\alpha=L,C,R$ labels the left, center, or right QPC respectively; $\gamma=1,2$ labels the electron localization on either side of the QPC; and $\epsilon_{\alpha\gamma k}$ is the dispersion of an electron with momentum $k$, with corresponding fermionic creation (annihilation) operators $c_{\alpha\gamma k}^{\dagger}$ ($c_{\alpha\gamma k}^{\phantom{\dagger}}$). Tunneling at each QPC is described by,
\begin{equation}\label{eq:Hqpc}
H_{\rm QPC} = \sum_{\alpha} J_{\alpha} \left ( c_{\alpha 1 }^{\dagger}c_{\alpha 2}^{\phantom{\dagger}} + {\rm H.c.} \right ) \;,
\end{equation}
where $c_{\alpha\gamma}=\sum_{k} a_{\alpha\gamma k} c_{\alpha\gamma k}$ are local orbitals at the QPC. While at first glance there is no apparent process which tunnels an electron between $\alpha = L,R \text{ and } C$, electrons have a long dwell time and decohere across the island, effectively making the two sides of each island independent channels. This is valid since the island hosts a large number of electrons with a small level spacing. Thus, electrons on the island can freely diffuse between $\alpha = L,R \text{ and } C$ allowing for series conductance. Note that the experimental transport measurement effectively maps to a spin current measurement in the model. An electron on the left lead which tunnels onto the left island flips the left island's pseudospin. Tunneling to the right island flips the left island's pseudospin back, while also flipping the right island's pseudospin. Finally, tunneling to the right lead again flips the right island's pseudospin back to its original state. The device configuration is then effectively `reset', allowing series transport of the next electron between the leads across the two-site structure.

Finally, electronic interactions are embodied by,
\begin{equation}\label{eq:Hint}
H_{\rm int} =  E_C^L \hat{N}^2 + E_C^R \hat{M}^2 + V \hat{N}\hat{M} \;,
\end{equation}
where $E_C^L$ and $E_C^R$ are the local charging energies of the left and right islands, while $V$ is the inter-island capacitive interaction. Here $\hat{N}=\sum_k (c_{L 2 k}^{\dagger}c_{L 2 k}^{\phantom{\dagger}}+c_{C 1 k}^{\dagger}c_{C 1 k}^{\phantom{\dagger}})$ and $\hat{M}=\sum_k (c_{R 2 k}^{\dagger}c_{R 2 k}^{\phantom{\dagger}}+c_{C 2 k}^{\dagger}c_{C 2 k}^{\phantom{\dagger}})$ are operators for the total number of electrons on the left or right islands. The occupancy of the islands can be tuned by applying plunger gate voltages $P_{L,R}$, which we convert to energies $U_{L,R}=U_{L,R}^0+\alpha P_{L,R}$ using the experimentally measurable capacitive lever arm $\alpha= 50~\mu\text{eV/mV}$, relative to the reference $U_{L,R}^0$. This corresponds to a term in the Hamiltonian $\hat{H}_{\rm gate} = B_L \hat{N} + B_R \hat{M}$ where $\vec{B}=\bar{\boldsymbol{\alpha}} \vec{U}$ accounts for cross-capacitive gate effects through the dimensionless 2x2 matrix $\bar{\boldsymbol{\alpha}}$. The total Hamiltonian describing the device is $\hat{H}=\hat{H}_{\rm elec} + \hat{H}_{\rm QPC} + \hat{H}_{\rm int} + \hat{H}_{\rm gate}$.

Following a similar analysis by Matveev~\cite{Matveev1991,Matveev1995} for the single-island case, we transform the model into a generalized quantum impurity model involving macroscopic charge pseudospins on the two islands. This assumes that one can separate the dynamics of the interacting reservoir of island electrons into the dynamics of an auxiliary charge operator and a free electron bath, with changes in the total charge of the islands kept track of through the QPC tunneling terms. The validity of this assumption has been stringently confirmed by the quantitative agreement between model predictions and recent experimental results~\cite{Iftikhar2015,Iftikhar2018,mitchell2016universality,han2021extracting}. We define island charge pseudospin operators $\hat{\mathcal{S}}_L^+ = \sum_N |N+1\rangle\langle N|$ and $\hat{\mathcal{S}}_R^+ =\sum_M |M+1\rangle\langle M|$, with $\hat{\mathcal{S}}_{L,R}^- = (\hat{\mathcal{S}}_{L,R}^+)^{\dagger}$. We label lead electrons $\alpha\gamma = L1$ and $R1$ as `up' spin and the island electrons $L2$ and $R2$ as `down' spin, such that $c_{L2}^{\dagger}c_{L1}^{\phantom{\dagger}} \equiv \hat{s}_L^-$ and $c_{R2}^{\dagger}c_{R1}^{\phantom{\dagger}} \equiv \hat{s}_R^-$, with $\hat{s}_{L,R}^+ = (\hat{s}_{L,R}^-)^{\dagger}$. We also relabel $C1\to C_L$ and $C2\to C_R$. Then Eq.~\ref{eq:Hqpc} becomes,
\begin{equation}\label{eq:Hqpc_spin}
\begin{split}
H_{\rm QPC} =& \left ( J_L^{\phantom{+}} \hat{\mathcal{S}}_L^+ \hat{s}_L^- + J_R^{\phantom{+}} \hat{\mathcal{S}}_R^+ \hat{s}_R^- + {\rm H.c.} \right ) \\
& + \left (J_C^{\phantom{+}} \hat{\mathcal{S}}_L^+ \hat{\mathcal{S}}_R^- c_{C_L}^{\dagger}c_{C_R}^{\phantom{\dagger}} + {\rm H.c.} \right ) \;.
\end{split}
\end{equation}
The first line of Eq.~\ref{eq:Hqpc_spin} corresponds to the transverse components of exchange interactions between each island's charge pseudospin and an effective spinful channel of conduction electrons. The second line is the novel ingredient in this system: a correlated tunneling at the central QPC, involving a mutual pseudospin flip on the islands. The form of this interaction follows naturally from the same mapping used in generating the first line. 

This full model, including multiple charge states for each island, is solved numerically-exactly using interleaved-NRG (iNRG)~\cite{mitchell2014generalized,stadler2016interleaved} to obtain the conductance plots in Figs.~\ref{fig:stabNRG20}, \ref{fig:stabNRG02} and \ref{fig:linecutcalc}. Further details are given in the Supplementary Info. 

Note that at low temperatures and small QPC transmissions, only the lowest two charge states of each island can be retained in the model, and the rest safely neglected. The operators $\hat{\mathcal{S}}_L$ and $\hat{\mathcal{S}}_R$ then become spin-$\tfrac{1}{2}$ operators, and we recover standard Kondo exchange interactions (albeit anisotropic). Furthermore, near the triple point, the universal quantum critical physics is captured by a reduced model involving only the three lowest degenerate charge states of the two-site structure,
\begin{equation}
\begin{split}
    H_{\rm TP} =H_{\rm elec}+  (&J_L \hat{s}_L^- |N+1,M\rangle\langle N,M| \\+ &J_R \hat{s}_R^- |N,M+1\rangle\langle N,M| \\+ &J_C \hat{s}_C^- |N+1,M\rangle\langle N,M+1| +{\rm H.c.}) \\
     + &\epsilon_L |N,M\rangle\langle N,M| \\+&\epsilon_R |N+1,M\rangle\langle N+1,M| \\+&\epsilon_C |N,M+1\rangle\langle N,M+1| \;,
\end{split}
\end{equation}
where $\hat{s}_{\alpha}^\pm$ are local spin-$\tfrac{1}{2}$ operators for independent conduction electron channels $\alpha=L,R,C$; the three retained double-island charge states are $|N,M\rangle$, $|N+1,M\rangle$, and $|N,M+1\rangle$; and $\epsilon_{\alpha}$ are the gate-tunable effective energies of these states. 

The Kondo frustration in this model is illustrated in Fig.~1b.
This reduced model for the critical point is used to obtain the universal results in Fig.~3a,b.


\subsection*{Device}
The device was fabricated on a GaAs/AlGaAs heterostructure with a 2DEG approximately 95nm deep, density of $2.6 \times 10^{11} \text{ cm}^{-2}$ and mobility $2.0 \times 10^6 \text{ cm}^2\text{V}^{-1}\text{s}^{-1}$. An SEM micrograph of an equivalent device is shown in Extended Data Fig.~\ref{fig:SEM}. High quality of the small ohmic contacts is crucial, so we take special steps to ensure cleanliness. Before any fabrication is done on the heterostructure, we dip in HCl $3.7\%$ to remove any oxide layer that has built up. After writing the ohmic layer pattern using e-beam lithography and developing the PMMA resist, we use a light oxygen plasma etch to remove residual PMMA scum. Next, before evaporating the ohmic stack, we use the following chemical treatment procedure: dip in TMAH $2.5\%$ for 20 seconds, 5 seconds in water, 5 seconds in HCl $37\%$, 5 seconds in water (separate from the first cup of water). Afterwards, we quickly move the chip into a KJL Lab 18 e-beam evaporator, and pump down to vacuum. Reducing the time in air is important to prevent substantial oxide layer growth. Finally, we run an in-situ Argon etch for 20 seconds. Only after this do we evaporate the ohmic stack (107.2 nm Au, 52.8 nm Ge, and 40 nm Ni, in order of deposition). 

\subsection*{Experimental Setup}
The device was cooled down with a $+300$ mV bias on all the gates to reduce charge instability by limiting the range of voltages we need to apply~\cite{Pioro-Ladriere2005}. To reduce thermoelectric noise causing unwanted voltage biasing across the device, each lead has a central ohmic contact (between the source and measurement contacts) which are all shorted to each other on chip. The shorted ohmics are connected to a single line and grounded at room temperature.  Measurements are made at low frequencies ($<100 \text{ Hz}$) using an SR830 lock-in amplifier. The 14 mV output of the SR830 is current biased using a 100M$\Omega$ resistor, and the current is then converted to a voltage on chip by the quantum Hall resistance ($h/2 e^2$). A measurement of either the series transmitted voltage or the reflected voltage is amplified by an NF SA-240F5 voltage amplifier. This voltage simply converts to the series conductance. For most reported measurements, we source at \textit{S2} and measure at \textit{M2} (see Fig.~\ref{fig:schematic}). The series conductance is then related to the reflected voltage, $V_2$ by Eq.~\ref{eq:Gseries}.
\begin{equation}
G = \frac{e^2}{3h} \frac{V_2 - V_2^{\tau_{R,L,C} =0}}{V_2^{\tau_{R,L,C} =1} - V_2^{\tau_{R,L,C} =0}}
\label{eq:Gseries}
\end{equation}
Measurements in this way eliminate the need for precise knowledge of many settings in a given setup -- excitation amplitude, amplifier gain, line resistances, etc. For arbitrary sourcing and measurement configurations, the relations between measured voltages and the desired conductances can be calculated straightforwardly through Landauer–Büttiker formalism. 


\subsection*{Electron Temperature}
The electron temperature is determined from dynamical Coulomb blockade measurements outlined in Iftikhar~\cite{Iftikhar2016}. The zero bias suppression of the conductance across a QPC when series coupled to another QPC of low resistance can be fit to a known theoretical from which directly depends on the electron temperature~\cite{Joyez1997}. In our device we measure through two QPCs across a single island ($\tau_C = 0$), with one QPC partially transmitting and the other set to fully transmit a single edge mode.

\subsection*{Calibrating QPC Transmissions} \label{sssec:qpcCalib}
A standard procedure to measure the transmission of each QPC is to measure the series conductance while varying the applied gate voltage of the QPC, with all other QPCs set to fully transmit $n$ edge states, acting as an $h/ne^2$ resistor. For our experiments of central interest, we must then adjust each QPC to a desired transmission. We cannot naively use the gate voltages that produced that transmission with all other QPCs open, because changing the voltage applied to any gate capacitively affects all other QPCs. However, we can calibrate this capacitive effect by measuring how much each QPC's transmission curve shifts as we vary each other QPC's gate voltage by a known amount. This is done for each QPC we use in the experiment, and with this information we can systematically determine the appropriate gate voltages to set. However, even this procedure fails in our device. When transmission is measured in series through a QPC and an additional resistance on order $h/e^2$, dynamical Coulomb blockade (DCB) suppresses the conductance relative to that expected from ohmic addition of the `intrinsic' transmissions~\cite{Souquet2013,Altimiras2007,Joyez1997,Senkpiel2020,Jezouin2013,Flensberg1992,Yeyati2001,Pekola2008,Golubev2004,Devoret1990,Parmentier2011}. The `intrinsic' transmissions can be recaptured by applying a source drain bias large compared to a relevant charging energy (Extended Data Fig.~\ref{fig:leadQPCcomp}). 

Alternatively, a measurement pathway which does not go through the metallic island (for example, measuring $\tau_R$ through the plunger gate $P_R$) effectively shorts the circuit to ground, so the `intrinsic' transmission is recovered even at zero source drain bias~\cite{Altimiras2007, Slobodeniuk2013}. The measured QPC transmission should then be the same as the measurement through the island at high bias. Empirically, this is not the case in our device. Repeating the measurement through $P_R$ at high bias shows exact agreement with the high bias measurement through the island. Bias dependent measurements through $P_R$ are entirely consistent with DCB suppression, contrary to expectation (Extended Data Fig.~\ref{fig:plgQPCcomp}). 

We suspect that this residual DCB effect is due to impedances in our measurement setup, external to the device. We connect our measurement lines which ground the device through highly resistive coaxial lines and discrete filters located right above the connection to the sample through large ohmic contacts. This is in contrast to earlier work (\cite{Iftikhar2015, Iftikhar2018}) on this type of system in which a cold ground is used, effectively creating a very low impedance path to ground. 

The important question is which measurement transmission is relevant for the Kondo interactions. While previous work~(\cite{Iftikhar2015, Iftikhar2018}) found the difference between the 'intrinsic' transmission and the zero bias transmission through an open plunger gate to be small, and thus both equally valid, we find that the measurement through $P_R$ at zero bias empirically works best in our device. Without any residual DCB, we believe this should be the exact same as the ‘intrinsic’ transmission. However, the residual DCB appears to not only suppress the measured QPC transmissions, but all measured conductances of our system in any configuration, at zero bias. In particular, when measuring through either one or both islands, the conductance when using the ‘intrinsic’ transmission at charge degeneracy points, where Kondo interactions are most important, is consistently lower than both expectations and previous results seen in \cite{Iftikhar2015}. With the single island, we see a suppression consistent with DCB -- the conductances away from 0 and .5 are suppressed the most, and the suppression is reduced at higher temperatures, where DCB is weaker. 

If instead we use not the ‘intrinsic’ transmission, but the zero bias transmissions through $P_R$, we see remarkably consistent agreement with previous results on two-channel Kondo for different transmissions and temperatures. This is shown and described in the Supplementary Info. Our interpretation is that while both Kondo and DCB-renormalizations occur, we can work in a ‘DCB-renormalized space’ by explicitly setting the DCB-renormalized transmissions. These DCB-renormalized transmissions then act as the transmissions that are important for the Kondo interactions. Importantly the transmissions must be set in the same space that DCB is renormalizing the measurements of interest. This means using the zero bias measurement through $P_R$, where DCB is caused by only the external impedances, and not through the island, where there is additional suppression due to the resistance of the second QPC in series. 

On the left island, the QPC we use does not have an adjacent pathway through $P_L$, but due to an equivalence of the island-lead QPCs, a consistent mapping can be made from the `intrinsic' transmission to the DCB-renormalized one that would be measured through the plunger gate. However, we are unable to make this mapping for the inter-island QPC as we observe differences in the DCB-renormalization when measured through the islands (Extended Data Fig.~\ref{fig:interislandQPCcomp}). Likewise,  we are unable to measure the extraneous DCB-renormalization of the inter-island QPC due to the device geometry. In our results in the main text, we therefore report `intrinsic' values for $\tau_C$. This may be why the maximum conductance does not appear at $\tau=\tau_C$, and in any case it means that the $\tau_C$ relevant for Kondo physics grows with increasing temperature.

Regardless of whether the transmission settings in the main text are reported as the `intrinsic' values or the DCB-renormalized values, the conclusions of the main text hold. None of our results rely on precise quantitative knowledge of the transmissions settings, since $T^*$ depends only on detuning of one transmission relative to another. This would not be the case in any future work exploring universal scaling as a function of $T/T_K$, since $T_K$ depends directly on the quantitative transmissions.

\subsection*{Fitting universal curves}
To convert $|\Delta U|^{3/2}$ to $T^*$ requires the unknown scaling prefactor $b$ of Eq.~\ref{eq:Tstar}. In Fig.~\ref{fig:tauCollapse}, we do this by least-squares fitting $|\Delta U|^{3/2}$ of the experimental data to $T^*$ of the approximately universal curve, with $b$ as a free parameter. Each $\tau_C$ line cut is independently fit, and we average the resulting $b$ values and apply the same rescaling to each curve. We find ${b=0.858}$~mK/${\mu\text{eV}}^{3/2}$ works best. A similar procedure is done in Fig.~\ref{fig:tempCollapse}, except we fit only the 20 mK data to the fully universal curve, obtaining ${b=0.769}$~mK/${\mu\text{eV}}^{3/2}$. The elevated temperature data are then rescaled with that same obtained $b$. For every curve, we also fit a constant shift $a|\Delta \tau_C|^{3/2}$ to take into account that there is a finite $T^*$ even at $\Delta U =0$ due to a detuning from the critical couplings. 

Because of the influence of neighboring triple points (TPs), we exclude the data at large $T^*$ in our fitting procedure, where there is a significant change in the slope. For the temperature data specifically, the first couple points at low $T^*$ are also excluded, due to the clear deviation from universality which we understand as a consequence of a finite $T/T_K$. While this is captured by an approximately universal curve as in Fig.~\ref{fig:tauCollapse}, the comparison fails when exploiting the universal parameter $T^*/T$. At each temperature $T/T_K$ also changes, and thus a different approximately universal curve would be needed. 
A clear limit to how high of temperatures we can observe collapse is when the TPs are extremely broadened such that the conductance is no longer reflecting exclusively the line shape of the TP. In fact, we already see signatures of this with the 75 mK data plotted in Fig.~\ref{fig:tempCollapse}, where the conductance falloff with $T^*/T$ is everywhere slower than in the universal conductance curve.

In Extended Data Fig.~\ref{fig:tauValues}, we show the $\tau_C$ values used in the scaling collapse data of Fig.~\ref{fig:tauCollapse} in the main text. In order to resolve the TP peaks we need a sufficiently large $\tau_C$, of which the value needed grows with $\tau$. To reach a conductance close to $e^2/3h$ and have split triple points, $\tau_C$ must be made extremely close to 1. While this may indicate the measured line cuts are essentially identical, we see that in the full shape (Extended Data Fig.~\ref{fig:highTauRaw}) there are large changes to the line cuts. We can compare this to Fig.~\ref{fig:linecut20} of the main text where large changes in $\tau_C$ are needed for comparable changes in the line cuts.

In Fig.~\ref{fig:tempCollapse}, the set transmissions at each temperature are not the same in order to roughly match the conductance at ${\Delta{U} = 0}$. If the transmissions are held constant, the conductance would decrease with increasing temperature due to a larger $T/T_K$. By increasing $\tau,\tau_C$ with temperature, $T/T_K$ is roughly constant. In our measurement, $\tau = \left\{0.78, 0.78, 0.81, 0.82\right\}$ at $T=\left\{20, 26, 46, 75\right\}$ mK while $\tau_C = 0.9$ for all $T$. As described in the Calibrating QPC Transmissions section of Methods, we know that this corresponds to a slightly increasing $\tau_C$ with temperature, although the exact value is unknown. That $\tau_C$ varies between temperatures does not affect the conclusion, since the scaling as a function of $\Delta U$ is unaffected. 

Finally, as stated in the main text, the TPs are not easily identifiable, depending on $\tau$ and $\tau_C$. While in Fig.~\ref{fig:scalingCollapse} we try to choose line cuts in which the TPs are relatively well separated, it is not always possible depending on the couplings and temperature. By using NRG calculations at extremely low temperatures to determine the TP locations and comparing them to the location of the conductance peaks at higher temperatures, we can estimate the difference between conductance peak and TP position. We find from our NRG calculations that choosing a point $\sim 1$ $\mu$eV away from the peak conductance for the two lowest temperatures of Fig.~\ref{fig:tempCollapse}, $\sim 2$ $\mu$eV away for the 46 mK data, and $\sim 10$ $\mu$eV away for the 75 mK data is a good approximation. The large offset of the 75 mK data is simply due to there only being a single conductance peak at the center between both TPs. For all of the data used in Fig.~\ref{fig:tauCollapse}, where the TPs are well separated, we find that using the conductance peak location as the TP location is a good approximation. Due to a combination of an imperfect approximation in identifying the TP location and measurement resolution, there will still be some error in choosing the true TP location. We estimate this uncertainty in $\Delta{U}$ to be $\sim 0.25$ $\mu$eV for all data taken at the two lowest temperatures, and $\sim 1$ $\mu$eV for the two higher temperatures. Though any uncertainty in $\Delta{U}$ creates a non-linear error when scaling $\Delta{U}$ to $T^*$, we find that changes within the uncertainty of where we set $\Delta{U} =0$ do not significantly change our results. We see this by offsetting $\Delta{U}$ and redoing the same scaling collapse analysis.
\subsection*{Alternative $T^*$ scaling}

A natural question is whether our experimental system could be realized by an alternative model. Demonstrating that an experimental system is a realization of a particular theoretical model is always complicated, since the space of possible alternative models and refinements is in principle infinite. That said, the universality classes that emerge from renormalization group analysis offer one way of distinguishing different classes of models and of identifying which could satisfactorily describe experimental data. We believe the scaling collapse to a universal curve using a non-trivial power law is remarkable and is the unique signature of the novel critical point studied. However, the experimental data alone are not sufficient to definitively distinguish the power law from another close one. It is the overall consistency between experiment and theory in all features which validates the novel model. Not only is our model the natural one for the system we’ve constructed, but we are aware of no alternative model whose predictions are consistent with all of our experimental results, as our NRG results are.

However, we can still do the same scaling collapse analysis assuming an alternative power law and observing the results. Explicitly, we assume $T^* = T^*_0 + b|\Delta U|^{\gamma}$, where $T^*_0 = a|\Delta \tau_C|^{\gamma}$, and plot the results in Extended Data Fig.~\ref{fig:scalingExp1} for $\gamma=1$ (other power laws such as $1/2$ or $2$ clearly do not work). As we outline in the previous section, we generally need to fit an unknown shift $T^*_0$ since we do not have direct knowledge of the deviation away from the critical couplings. When $\gamma = 1$ instead of $3/2$, there is no physically-motivated universal curve, so instead of finding $T^*_0$ by best fit to a universal curve, we apply a shift $T^*_0=a|\Delta \tau_C|$ such that $G=1/2 \text{ } e^2/3h$ at $T^*=T$ for the 20 mK data, the condition which defines the relative scale of $T^*$. The higher temperature data also need an independent $T^*_0$ shift, as the couplings are changed with temperature to maintain similar conductance at the peak, and thus $T^*_0$ also changes. We choose a $T^*_0$ for the higher temperatures such that the data best collapse onto the 20 mK data. 

We see that the experimental data are plausibly consistent with the power law of 1, but there are a few subtle deviations. For one, we would normally expect that deviations from collapse grow with increasing temperature, as the difference between scaled data based on actual vs. assigned power law exponent becomes more and more significant. We observe a clear deviation in the 46 mK data, making the seemingly good collapse of the 75 mK data paradoxically a sign of a worse fit. We can understand this by looking at the raw data from which the line cut is extracted in Extended Data Fig.~\ref{fig:raw75mk}. It is clear that at an elevated temperature, the two triple points are completely merged together. Since the universal curve captures the conductance of an isolated triple point, we expect that in the experiment the conductance at a particular value of $\Delta U$ would be higher than theoretically predicted. Thus, if we were to scale by the true exponent, the 75 mK data would overshoot the isolated conductance and not collapse as well to the lower temperature data. We see this happen when using an exponent of 3/2, whereas with an exponent of 1, the 75 mK data actually appears to collapse well. The same argument can be used for the poorer collapse of the 46 mK data. For larger $\Delta U$ (and thus $T^*$), the 46 mK line cut is clearly beneath the rest of the data. 

We see this in an alternative analysis by exploiting that $T^*=T$ when $G=1/2 \text{ } e^2/3h$. We can thus plot $T^*-T^*_0$ against $|\Delta U|$ and fit a power law to it. To do this, we first identify the $\Delta U$ values in which $G=1/2 \text{ } e^2/3h$ at each temperature. Since the 75 mK data never reach a conductance that low, we do not include it in our analysis. We note that a similar analysis done in previous work (\cite{Iftikhar2018}) only used data at their lowest temperatures, and saw deviations from the expected power law at higher temperatures. We can use the already determined $T^*_0$, but since these shifts are different for $\gamma=3/2$ and $1$ , we will show the results using both (Extended Data Fig.~\ref{fig:gammaFitExp32},~\ref{fig:gammaFitExp1} respectively). That we are not simply assuming what we are trying to prove is illustrated by the fact that in both analyses, $\gamma = 3/2$ is a better description than $\gamma = 1$ of the change between the two lower temperatures and 46 mK. This is a reflection of the poor scaling collapse of the 46 mK data when $\gamma = 1$ we saw in Fig.~\ref{fig:scalingExp1}. 

While the above analysis suggests $\gamma=3/2$, we acknowledge that definitively determining the relevant power law from experiment alone would require a larger range of temperatures. Since there is a clear practical high-temperature limit (when the triple points merge), obtaining a larger temperature range would require reaching even lower temperatures or larger charging energies. Again, it is the consistency of experimental data with the 3/2 power law combined with a broader set of agreement with NRG results that gives confidence that we are indeed observing the DCK model. 


\renewcommand{\figurename}{Extended Data Fig.}
\setcounter{figure}{0}
\begin{figure*}[h]
\centering
\includegraphics[width=3.2 in]{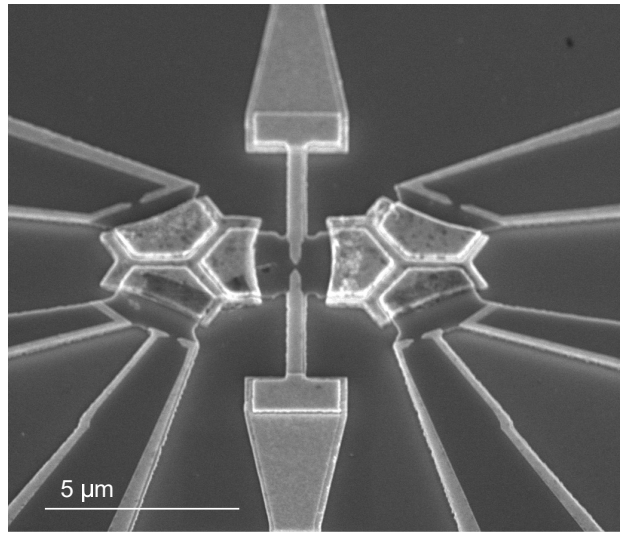}
\caption{\textbf{SEM micrograph of nominally identical device.}}
\label{fig:SEM}
\end{figure*}

\begin{figure*}[h]
    \centering
    \begin{subfigure}[h]{\linewidth} 
        \includegraphics[width=6.5 in]{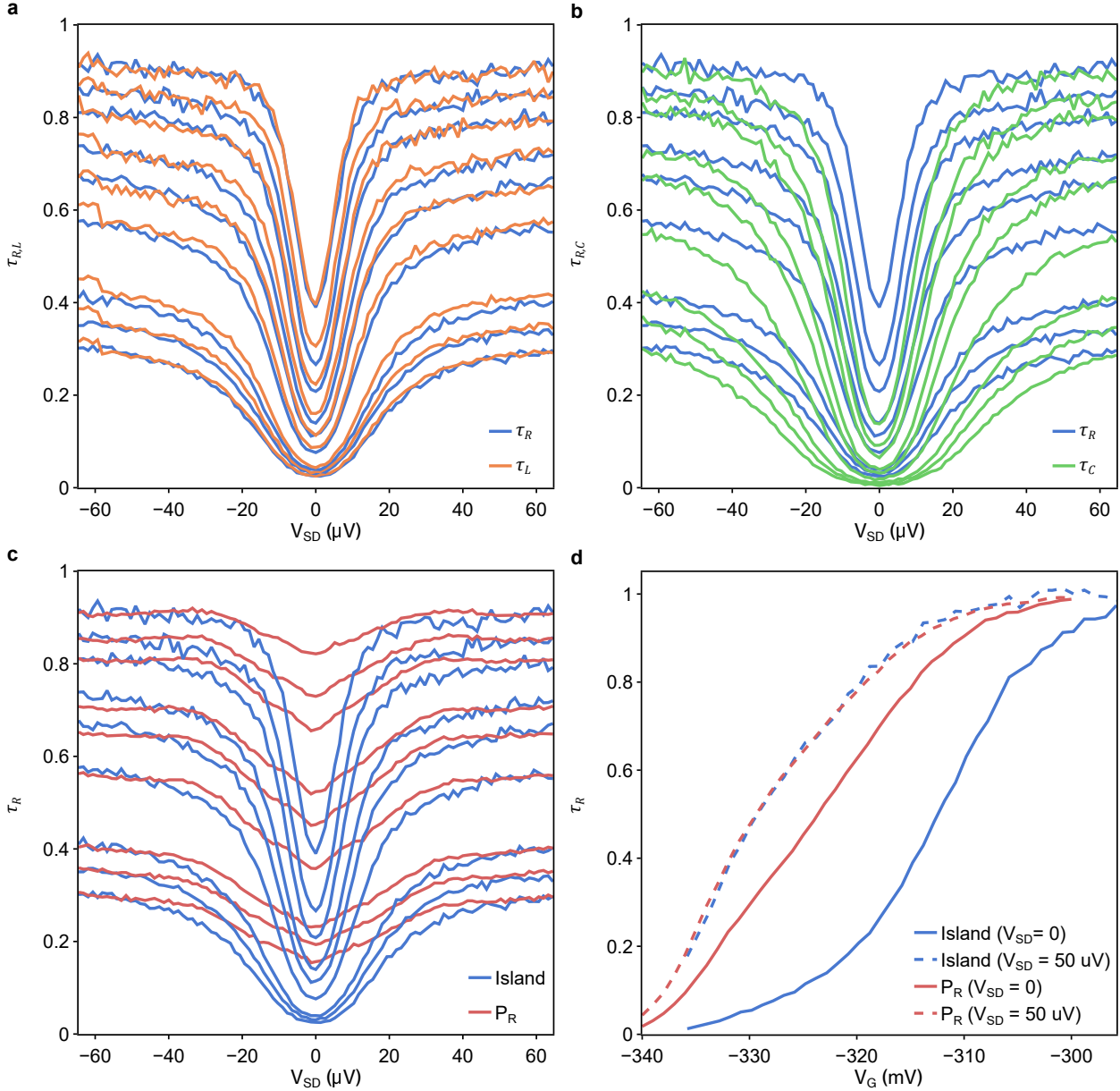}
        \phantomcaption
        \label{fig:leadQPCcomp}
    \end{subfigure}
    \begin{subfigure}[t]{0\textwidth} 
         \includegraphics[width=\textwidth]{RoughDraftFigures/figSuppQPCCalib.eps}
         \phantomcaption
         \label{fig:interislandQPCcomp}   
    \end{subfigure}
    \begin{subfigure}[t]{0\textwidth} 
         \includegraphics[width=\textwidth]{RoughDraftFigures/figSuppQPCCalib.eps}
         \phantomcaption
         \label{fig:plgQPCcomp}   
    \end{subfigure}
    \begin{subfigure}[t]{0\textwidth} 
         \includegraphics[width=\textwidth]{RoughDraftFigures/figSuppQPCCalib.eps}
         \phantomcaption
         \label{fig:biasQPCcomp}   
    \end{subfigure}
    \vspace{-10pt}
\caption{\textbf{Dynamical Coulomb blockade of QPC transmissions.} \textbf{a.} Measured QPC transmissions $\tau_R, \tau_L$ as a function of a source-drain bias $\text{V}_{\text{SD}}$ for different QPC gate voltages. The measured transmission is extracted by measuring the series conductance when in series with the inter-island QPC and opposite island-lead QPC set to fully transmit a single channel ($\tau_{C, L/R} = 1$). For the island-lead QPCs, this is equivalent to a measurement through a single island and the normally unused QPC set to a transmission of 1 (data not shown). The measured transmissions clearly dip at zero bias, consistent with dynamical Coulomb blockade (DCB) behavior. The high bias behavior ($\text{V}_{\text{SD}} \approx 50 \text{ uV}$) recovers the `intrinsic' transmission of the QPC, unrenormalized by DCB. \textbf{b.} DCB measurements comparing the right island-lead QPC to the inter-island QPC. It is clear there is a substantial difference in the DCB-renormalization at zero bias between the two, likely due to the device geometry. \textbf{c.} Comparison of measuring $\tau_R$ through both islands (blue lines, as in \textbf{a, b}) and through the adjacent plunger gate $P_R$ (red lines). While typically we would expect no significant bias dependence when measuring through $P_R$, we in fact see DCB-like behavior. \textbf{d.} Comparing the two measurement pathways of \textbf{c} at fixed source-drain bias as a function of the QPC gate voltage. The `through the island' (blue) measurements have been shifted by $9 \text{ mV}$ to account for the large capacitive cross-talk effect when switching between the two different measurement pathways. That the high bias traces match well is indicative that there is indeed DCB-renormalization of the transmission when measuring through $P_R$. Empirically, using the zero bias, `through the plunger gate' measurement of the transmission (solid red line), best captures the relevant transmissions in the Kondo interactions of our system.}
\label{fig:QPCCalib}
\end{figure*}

\begin{figure*}[h]
\centering
    \begin{subfigure}{\columnwidth}
        \centering
        \includegraphics[width=3.2 in]{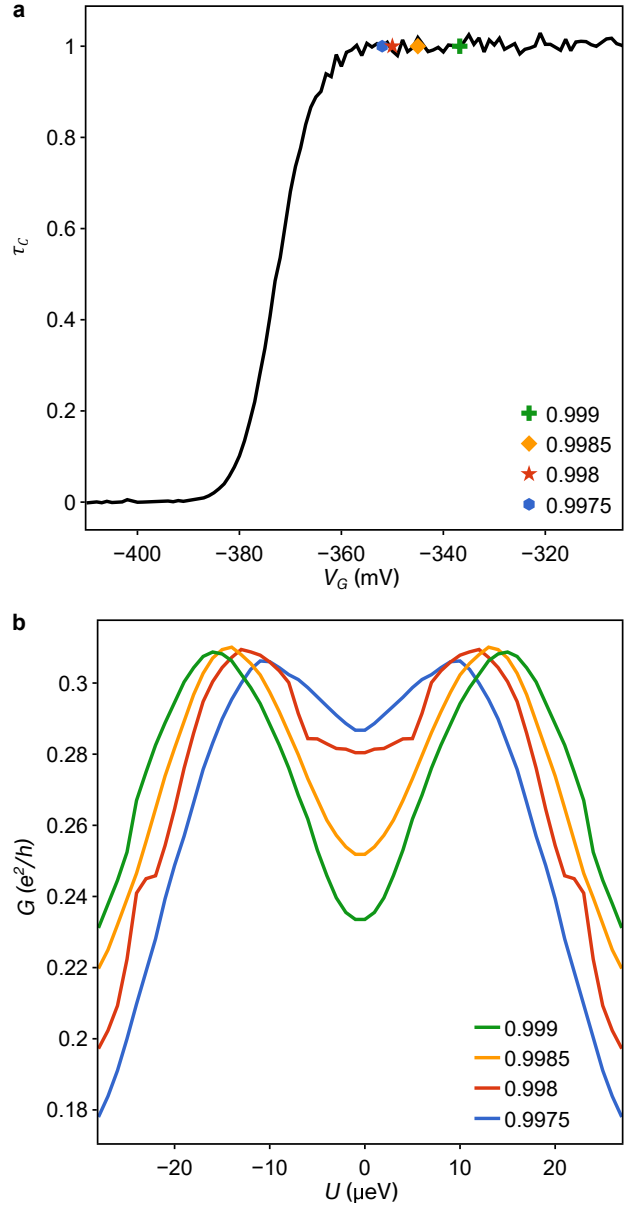}
        \phantomcaption
        \label{fig:tauValues}
    \end{subfigure}\ignorespaces
    \begin{subfigure}{0\columnwidth}
        \centering
        \includegraphics[width=\columnwidth]{RoughDraftFigures/figSuppTauCurves.eps}
        \phantomcaption
        \label{fig:highTauRaw}
    \end{subfigure}\ignorespaces
\caption{\textbf{Semi-universal $\tau_C$ values.} \textbf{a,} Measured inter-island transmission as a function of an applied gate voltage. The markers correspond to the inter-island transmissions used in Fig.~\ref{fig:tauCollapse} of the main text. \textbf{b,} Original line cuts in which the truncated data used in Fig.~\ref{fig:tauCollapse} are extracted from.}
\label{fig:highTauSupplement}
\end{figure*}

\begin{figure*}[h]
    \centering
    \begin{subfigure}[h]{\linewidth} 
        \includegraphics[width=6.5 in]{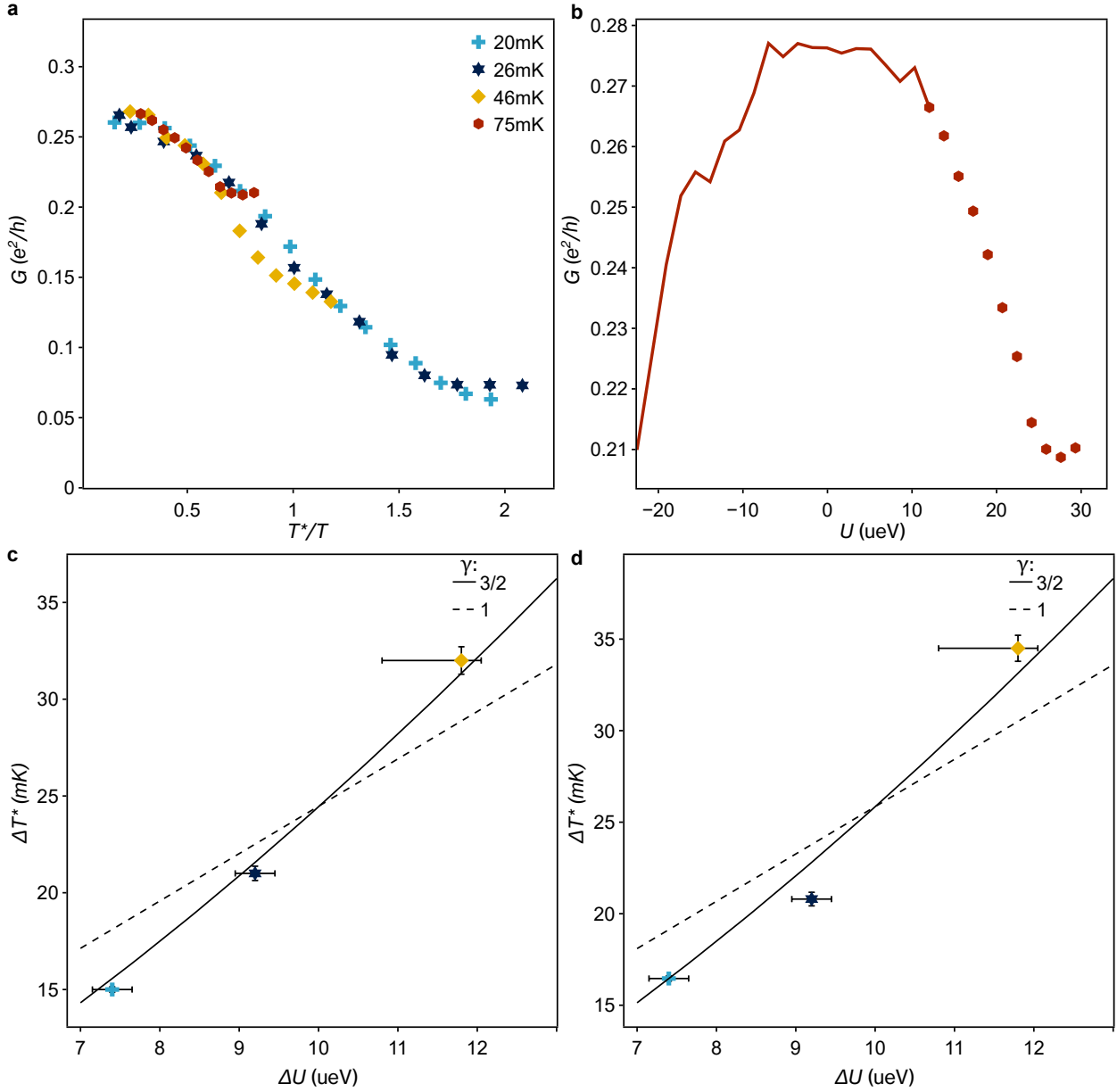}
        \phantomcaption
        \label{fig:scalingExp1}
    \end{subfigure}
    \begin{subfigure}[t]{0\textwidth} 
         \includegraphics[width=\textwidth]{RoughDraftFigures/figSuppScalingExponent.eps}
         \phantomcaption
         \label{fig:raw75mk}   
    \end{subfigure}
    \begin{subfigure}[t]{0\textwidth} 
         \includegraphics[width=\textwidth]{RoughDraftFigures/figSuppScalingExponent.eps}
         \phantomcaption
         \label{fig:gammaFitExp32}   
    \end{subfigure}
    \begin{subfigure}[t]{0\textwidth} 
         \includegraphics[width=\textwidth]{RoughDraftFigures/figSuppScalingExponent.eps}
         \phantomcaption
         \label{fig:gammaFitExp1}   
    \end{subfigure}
    \vspace{-10pt}
\caption{\textbf{Analysis of alternative power law.} \textbf{a.} Scaling collapse of the same data in Fig.~\ref{fig:tempCollapse} except here assuming ${T^*=T^*_0+b|\Delta U|^\gamma}$ where $\gamma = 1$. Other reasonable choices for $\gamma$ such as $1/2$ or $2$ do not show clear collapse onto a single curve. \textbf{b.} The original line cut between triple points from which the 75 mK data is extracted from, demonstrating the significant overlap of the two triple point peaks. \textbf{c, d.} For the three lowest temperatures, we plot $\Delta T^*=T^*-T^*_0$ against the $\Delta U$ value for which the conductance is $\frac{1}{2} e^2/3h$ and thus $T^* \approx T$. Since we determine $T^*_0$ from our scaling collapse analysis, it is necessarily dependent on the assumed power law. In \textbf{c} (\textbf{d}) we plot the $\Delta T^*$ values for which $\gamma=3/2 \text{ } (1)$ is assumed. The solid and dashed black lines are $\Delta T^*=b\left|\Delta U\right|$ power laws for $\gamma =3/2$ and $1$.}
\label{fig:gammaAnalysis}
\end{figure*}
\clearpage
\section*{Supplementary Information}

\subsection*{Metallic Island Interface}
To apply our model to our experimental system, the edge mode transmitted through a quantum point contact must then be perfectly transmitted into the metallic island. First, the edge mode must not bypass the metallic island. To enforce this, trenches are etched in the 2DEG below the island so the metal is the only conducting path bridging otherwise separate regions of 2DEG on different mesas. Second, reflection of the edge mode from the metal must be minimal. We verify near-perfect transmission by relating measured voltages to the transmission into the metal $\tau_{\Omega}^{R,L}$. In principle it is possible to extract the separate transmission probabilities for each metal-2DEG interface, but here we report an average. Unlike in the work of Iftikhar~(\cite{Iftikhar2015}), we find we must take into account the non-negligible resistance to ground in our setup -- through cryostat lines and filters intended to maintain low electron temperature -- which generates a finite transmitted voltage $V_T^{\tau_i=0}$, even with all QPCs closed. 
\begin{equation}
    \frac{V_{2(1)}^{\tau_{R(L), 4(3)} =1, \tau_C =0}-V_T^{\tau_i=0}}{V_{2(1)}^{\tau_{i} =0}-V_T^{\tau_i=0}} = 1-\tau_{\Omega}^{R(L)}/4
\end{equation}
Here $\tau_{3,4}$ are the unused QPCs in our experiment, representing the transmissions of the bottom QPCs of the left and right islands respectively. In our device, we measure
\begin{align*}
    \tau_\Omega^{R} = 1.0004 \pm 0.0098 \qquad \tau_\Omega^{L} = 0.9947 \pm 0.0417
\end{align*}
There is a much larger uncertainty on the left island, due to a noisier measurement contact $M1$.


\subsection*{Measurement Uncertainty}
A few factors contribute to uncertainty in the reported conductance. Due to slow drift in the electric potential, we cannot reliably sit at the same position in the charge stability diagram. This reduces the ability to average over time at a particular configuration of $P_L, P_R$. Based on the standard deviation in conductance in a configuration where each QPC was fully open, this error is {${\sim}0.001 \text{ } e^2/h$}. 

There is a second uncertainty in that the conductance for a given pair of triple points varies between periods in $P_L, P_R$. We observe that repeated measurements center around a particular value, with a few clear outliers of much lower conductance. To reduce this effect, we take the median conductance over a few periods. Reducing the error is limited by needing to remeasure the full charge stability diagram multiple times. For example, the data in Fig.~\ref{fig:linecut20} is extracted from 141 charge stability diagrams, of which about only 40 had no clear charge noise along the line between triple points. The errors of Fig.~\ref{fig:linecut20} for a particular $\tau_C$ and $U$ (median absolute deviation {${\sim}0.006 \text{ } e^2/h$}) are such that the exact $\tau_C$ line cut in which the conductance peaks at $U=0$ is unknown, but the overall non-monotonic behavior is unchanged. 

While not very frequent, charge noise may also shift the set transmission of one of the QPCs. Thus, for each 2D sweep we recalibrate each QPC to ensure we are at the transmission stated. The QPC calibration procedure relies on the accurate determination of the cross-talk between QPCs. While in principle this can be done very accurately, we cannot explicitly verify the transmissions in the experimental configurations used in our main results. However, we can verify that the procedure works by looking at the symmetry of the charge stability diagrams, since the two island-lead QPCs are made to have equal transmissions. From this we estimate that the transmissions are off by at most $0.03$.

Finally, our conversion from voltage to conductance could introduce errors. However, due to our ability to directly tune from $0$ to $1 \text{ } e^2/h$ without needing knowledge of the parameters of our measurement setup, this error is minimized. Any error is likely due to any imperfections in the metal-2DEG interface, as this would change the series conductance. Using our measurements of $\tau_\Omega$ above, we estimate that this error is $\sim 1\%$, coming from $\left| 1-\tau_\Omega^L\right| \sim .01$.

\subsection*{Orthogonal Line Cuts}
\begin{figure*}[h]
\centering
\includegraphics[width=3.2 in]{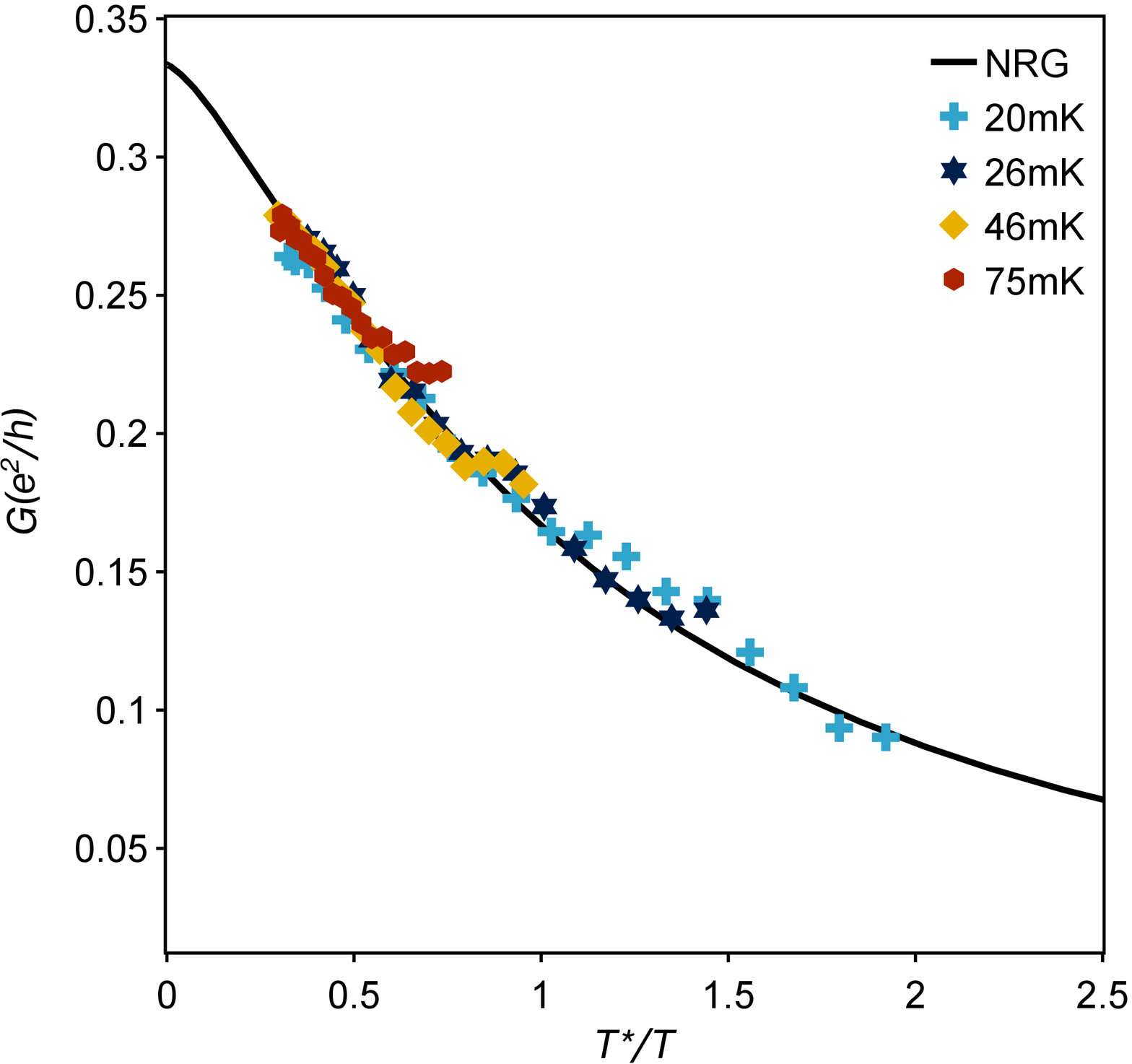}
\caption{\textbf{Orthogonal line cut scaling.} Scaling collapse for line cuts along an axis orthogonal to the triple point pair, using the same measurements from which the data in Fig.~\ref{fig:tempCollapse} is extracted. Similarly to the rescaling procedure used in the main text, the 20 mK data is scaled to match NRG data, while the plunger gate voltages for the line cuts at higher temperatures are scaled only by a $1/T^{2/3}$ factor.}
\label{fig:orthogScaling}
\end{figure*}

While the main text focuses on line cuts along the triple points, we emphasize that any detuning from the critical point should show the same power law scaling, due to the high symmetry at the triple point. For example, we can measure the conductance for a detuning in an orthogonal direction from the triple point axis, as shown in Extended Data Fig.~\ref{fig:orthogScaling}. Line cuts in this direction show the same $T^*$ scaling as along the triple point axis. However, interestingly, the unknown prefactor in the conversion from $\Delta{U} \rightarrow T^*$ is different. We find that the ratio of parallel to orthogonal prefactors is about $2.4$. 

\subsection*{Competition between Kondo interactions at $U=0$}
\begin{figure*}[h]
\centering
\includegraphics[width=3.2 in]{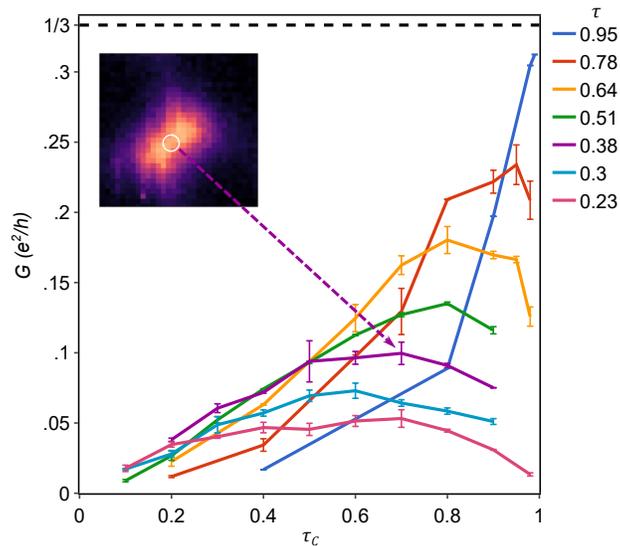}
\caption{\textbf{Competition between Kondo interactions.} Measured series conductance at the midpoint between the two triple points, for various $\tau_L=\tau_R=\tau$, plotted against $\tau_C$ at 20 mK. Inset: A representative charge stability diagram for $\tau=.38, \tau_C =.7$ from which the $U=0$ conductance is extracted.}
\label{fig:btwnTps}
\end{figure*}

In Extended Data Fig.~\ref{fig:btwnTps}, we show how the conductance changes when varying both $\tau$ and $\tau_C$. We focus here on the midpoint between triple points ($U=0$), which is immediately identifiable as a high symmetry point in the experimental gate voltage maps for all $\tau$ and $\tau_C$. We observe the same clear non-monotonic behavior as a function of $\tau_C$ seen in the line cuts of Fig.~\ref{fig:linecut20comp}, but here for a whole range of $\tau$. As $\tau$ is increased, the peak in conductance also occurs for larger $\tau_C$, reflecting that as the island-lead interaction strength is increased, the inter-island interaction also needs to be made stronger to maintain criticality. Finally, the peak conductance values also grow as $\tau$ is increased, which we can understand as an overall increase in the Kondo energy scale $T_K$. Only for the largest $\tau, \tau_C$, where $T/T_K \ll 1$, does the conductance approach $e^2/3h$. 

While we observe non-monotonic conductance as $\tau_C$ is detuned from $\tau_C^*$, with $\tau$ kept constant, the conductance should equivalently be non-monotonic as a function of $\tau$ if $\tau_C$ is kept constant. This is captured in Eq.~\ref{eq:Tstar}, where $\Delta \tau_C = \tau_C - \tau_C^*$, since $\tau_C^*$ is dependent on $\tau$. Thus, varying $\tau$ should also vary $T^*$ and lead to non-monotonic behavior. It is clear in Extended Data Fig.~\ref{fig:btwnTps} that for a given $\tau_C$, the largest conductance does not always correspond to the measurements with the largest $\tau$: conductance is enhanced when the different Kondo interactions are of comparable strength (so $T^*$ is minimized); contrariwise, tuning either $\tau$ or $\tau_C$ such that one Kondo interaction is stronger than another ($T^*$ is large) suppresses the series conductance. 


\subsection*{Single Island Two-Channel Kondo}
\label{sssec: 2CKComp}
\begin{figure*}[h]
\centering
\includegraphics[width=3.2 in]{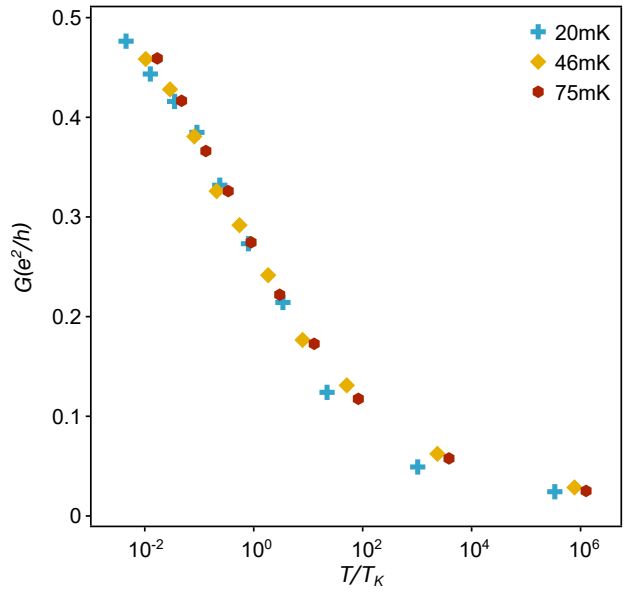}
\caption{\textbf{Two-channel Kondo scaling for a single island.} The measured series conductance through a single island ($\tau_C=0$), with equal transmissions set for the top and bottom island-lead QPC. The transmissions are rescaled into a Kondo temperature, and when the conductance is plotted as a function of the single parameter $T/T_K$, they fall onto a single universal curve.}
\label{fig:twoCKScaling}
\end{figure*}

A key assumption in this work is that both the left and the right island are true implementations of a metallic island. We reproduced a key result which demonstrate two-channel Kondo behavior shown by Iftikhar~\cite{Iftikhar2015}, which is only possible with a true equivalent recreation of the hybrid metal-semiconductor island. Shown in Extended Data Fig.~\ref{fig:twoCKScaling} is the series conductance through a single island ($\tau_C = 0$), rescaled into a universal conductance curve as a function of $T/T_K$. $T_K$ is determined by the set equal transmissions of two island-lead QPCs (for example, $\tau_R$, and the opposite QPC of the right island, unlabeled in Fig.~\ref{fig:schematic}). In practice, to best compare results, we set the same transmissions ($\tau = \left\{.06, 0.12, 0.245, 0.36, 0.47, 0.57, 0.68, 0.77, 0.85, 0.93\right\})$ as in \cite{Iftikhar2015}, and thus the same $T_K$ values (due to the similar charging energies $E_C$). A caveat in our system however, is that we could only see quantitative agreement when using the dynamical Coulomb blockade renormalized QPC transmissions described in Methods. 


\subsection*{Details of NRG calculations}
To simulate the experimental two island charge-Kondo device using NRG, the full model developed in the main text must be modified. In particular, only a finite number of charge states on each island can be retained in the calculations in practice. Using the pseudospin mapping, we write $H_{\rm NRG} = H_{\rm elec} + H_{\rm QPC} + H_{\rm int} +H_{\rm gate}$. Here,
\begin{equation}\label{eq:Helec_nrg}
H_{\rm elec} = \sum_{\alpha=L,R} \sum_{\sigma=\uparrow,\downarrow} \sum_k \epsilon_{k}^{\phantom{\dagger}} c_{\alpha\sigma k}^{\dagger}c_{\alpha\sigma k}^{\phantom{\dagger}} + \sum_{\gamma=1,2}\sum_k \epsilon_{k}^{\phantom{\dagger}} c_{C \gamma k}^{\dagger}c_{C \gamma k}^{\phantom{\dagger}} \;,
\end{equation}
describes two effective continuum conduction electron channels $\alpha=L,R$ with spin $\sigma=\uparrow,\downarrow$; and one spinless central electronic reservoir with flavors $\gamma=1,2$. We assume here for simplicity that the dispersions $\epsilon_k$ are equal for all channels, and correspond to a flat density of conduction electron states $\nu = 1/2D$ inside a band of half-width $D$. Tunneling at the QPCs give rise to effective pseudospin-flip terms described by,
\begin{equation}\label{eq:Hqpc_spin_nrg}
\begin{split}
H_{\rm QPC} =& \left ( J_L^{\phantom{+}} \hat{\mathcal{S}}_L^+ \hat{s}_L^- + J_R^{\phantom{+}} \hat{\mathcal{S}}_R^+ \hat{s}_R^- + {\rm H.c.} \right ) \\
& + \left (J_C^{\phantom{+}} \hat{\mathcal{S}}_L^+ \hat{\mathcal{S}}_R^- c_{C1}^{\dagger}c_{C2}^{\phantom{\dagger}} + {\rm H.c.} \right ) \;,
\end{split}
\end{equation}
where $\hat{s}_{\alpha}^-=c_{\alpha \downarrow}^{\dagger}c_{\alpha \uparrow}^{\phantom{\dagger}}$ and $\hat{s}_{\alpha}^+=(\hat{s}_{\alpha}^-)^{\dagger}$ are local conduction electron spin operators, with $c_{\alpha \sigma} = \sum_k a_k c_{\alpha \sigma k}$ and $c_{C \gamma}=\sum_k a_k c_{C \gamma k}$ localized orbitals at the QPC positions. Within NRG, the charge pseudospin operators $\mathcal{\hat{S}}_{L,R}$ for the left and right islands are defined as $\mathcal{\hat{S}}_L^+ = \sum_{N=N_0-\bar{N}}^{N_0+\bar{N}} |N+1\rangle_L{}_L\langle N |$ 
and $\mathcal{\hat{S}}_R^+ = \sum_{M=M_0-\bar{M}}^{M_0+\bar{M}} |M+1\rangle_R{}_R\langle M |$, with $\mathcal{\hat{S}}_{L,R}^-=(\mathcal{\hat{S}}_{L,R}^+)^{\dagger}$. Here $N_0$ and $M_0$ are reference fillings, while $\bar{N}$ and $\bar{M}$ determine the number of accessible island charge states. Formally $N_0,\bar{N},M_0,\bar{M}\rightarrow \infty$ in the thermodynamic limit, but in practice a finite number of charge states $\bar{N}$ and $\bar{M}$ can be retained in the NRG calculations -- provided the QPC transmission is not too high, and the temperature is low enough compared with the charging energies. One can check \emph{post hoc} that the results of NRG calculations are converged with respect to increasing $\bar{N}$ and $\bar{M}$ for a given set of physical model parameters.

Electron interactions are embodied by the charging terms,
\begin{equation}\label{eq:Hint_NRG}
\begin{split}
H_{\rm int}=&E_C^L (\hat{N} - N_0-\tfrac{1}{2})^2 + E_C^R (\hat{M}-M_0-\tfrac{1}{2})^2 \\&+ V (\hat{N}-N_0-\tfrac{1}{2})(\hat{M}-M_0-\tfrac{1}{2}) \;,
\end{split}
\end{equation}
with $\hat{N}=\sum_{k}c_{L\downarrow k }^{\dagger}c_{L \downarrow k}^{\phantom{\dagger}}+c_{C1 k }^{\dagger}c_{C1k}^{\phantom{\dagger}} $ the total number operator for the left island, and
$\hat{M}=\sum_{k}c_{R \downarrow k }^{\dagger}c_{R \downarrow k}^{\phantom{\dagger}}+c_{C2 k }^{\dagger}c_{C2 k}^{\phantom{\dagger}} $ the total number operator for the right island, such that $\hat{N}|N\rangle_L = N |N\rangle_L$ and $\hat{M}|M\rangle_R=M|M\rangle_R$. The island occupancy is manipulated \emph{in situ} with gate voltages. In the model employed in NRG, this enters as the following pseudo-Zeeman field term in the Hamiltonian,
\begin{equation}\label{eq:Hgate_nrg}
H_{\rm gate}= B_L \hat{N} + B_R \hat{M} \;.
\end{equation}
Eqs.~\ref{eq:Hint_NRG} and \ref{eq:Hgate_nrg} are defined such that $B_L=B_R=0$ corresponds to the charge degeneracy point for the isolated islands, and the high-symmetry point between triple points in the charge stability diagram of the full system. 

Note that for $E_C^{L,R}\to \infty$, only the lowest two charge states of each island survive, and the model reduces to a variant of the standard spin-$\tfrac{1}{2}$ two-impurity Kondo model, here with anisotropic exchange, but with the dynamics of an additional conduction electron bath in the central region correlated to the inter-impurity interaction -- a kind of two-impurity, three channel model. Separate island-lead Kondo effects now compete with an inter-island Kondo effect, giving rise to an entirely new quantum phase transition.This version of the model is used to obtain results in the universal regime, presented in Fig.~\ref{fig:scalingCollapse}. However, for the quantitative simulation of the real device shown in Figs.~\ref{fig:device} and \ref{fig:linecut20comp}, we retain finite charging energies and keep multiple island charge states.

The above model is solved using Wilson's NRG method~\cite{wilson1975renormalization,bulla2008numerical}, which involves discretizing the conduction electron Hamiltonian $H_{\rm elec}$ logarithmically, mapping to Wilson chains, and diagonalizing the discretized model iteratively. $N_s$ of the lowest energy states are retained at each step, resulting in an RG procedure which reveals the physics on progressively lower energy scales. 

Standard NRG cannot be used in this case, however, due to the complexity of the model at hand, with 6 spinless conduction electron channels. Instead, we use the `interleaved NRG' (iNRG) method~\cite{mitchell2014generalized,stadler2016interleaved}, which involves mapping $H_{\rm elec}$ to a single generalized Wilson chain. This dramatically lowers the computational cost of such calculations, and brings the numerical solution of the model within reach. For all iNRG calculations presented in this work, we use a logarithmic discretization parameter $\Lambda=4$, retain $N_s=35000$ states at each iteration, and exploit all abelian quantum numbers.

The experimental quantity of interest is the series dc linear response differential conductance,
\begin{equation}\label{eq:cond}
G_{\alpha\beta}=\frac{dI}{dV_b}\Bigg|_{V_b\to 0}
\end{equation}
where we take $I=-e\langle \dot{N}_{R\uparrow}\rangle$ to be the current into the right lead due to a voltage $V_b$ applied to the left lead. Here $\dot{N}_{R\uparrow}=\tfrac{d}{dt}\hat{N}_{R\uparrow}$ and $\hat{N}_{\alpha\uparrow}=\sum_{k}c_{\alpha \uparrow k}^{\dagger}c_{\alpha \uparrow k}^{\phantom{\dagger}}$. An ac voltage bias on the left lead can be incorporated by a source term in the Hamiltonian, $H_{\rm bias}=-eV_b\cos(\omega t) \hat{N}_{L \uparrow}$, where $\omega$ is the ac driving frequency. The dc limit is obtained as $\omega \to 0$. 

The geometry of the device means that the conductance cannot be related to a spectral function. Instead we use the Kubo formula~\cite{Galpin2014},
\begin{equation}\label{eq:kubo}
    G = \frac{e^2}{h}\lim_{\omega \to 0} \frac{-2\pi  ~{\rm Im}K(\omega)}{ \omega} \;,
\end{equation}
where $K(\omega)=\langle\langle \dot{N}_{L\uparrow} ; \dot{N}_{R\uparrow}\rangle\rangle$ is the Fourier transform of the retarded current-current correlator $K(t)=-i\theta(t)\langle [\dot{N}_{L\uparrow},\dot{N}_{R\uparrow}(t) ] \rangle$. Within iNRG, ${\rm Im}K(\omega)$ may be obtained from its Lehmann representation using the full density matrix technique \cite{weichselbaum2007sum}. The numerical evaluation is substantially improved by utilizing the identity ${\rm Im}K(\omega)=-\omega^2 {\rm Im}\langle\langle \hat{N}_{L\uparrow} ; \hat{N}_{R\uparrow}\rangle\rangle$ \cite{transport}. 

We use iNRG to calculate the conductance through the device from Eq.~\ref{eq:kubo} at a given temperature $T$, as a function of $B_L$ and $B_R$.

\subsubsection*{Fitting and model parameters}
\begin{figure*}[h]
\centering
\includegraphics[width=3.2 in]{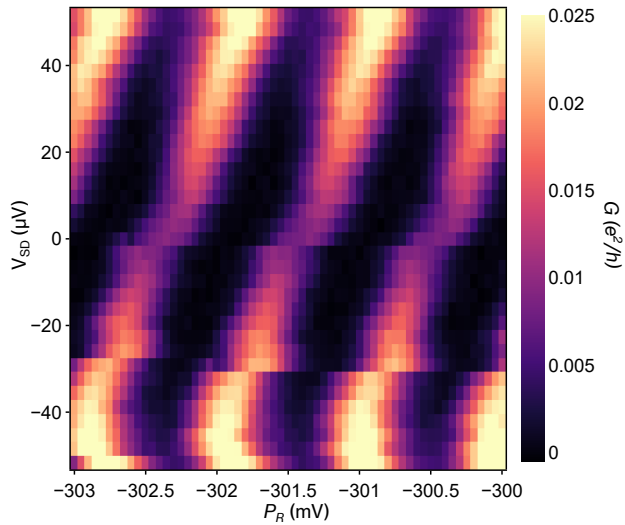}
\caption{\textbf{Coulomb diamonds.} Conductance measurement through a single island ($\tau_C = 0$) as a function of the plunger gate $P_R$ and a source-drain bias $V_{SD}$. The height of the diamond is used to extract $E_C \approx 25~\mu\text{eV}$, and subsequently a lever arm ${\alpha=50}~\mu\text{eV/mV}$ to convert gate voltages to energies.}
\label{fig:diamond}
\end{figure*}

Analysis of the experimental Coulomb diamond measurements (Extended Data Fig.~~\ref{fig:diamond}) yields $E_C^L\simeq E_C^R \simeq 25~\mu\text{eV}$ and the lever arm ${\alpha=50}~\mu\text{eV/mV}$ (which is assumed to be independent of other device parameters and temperature). We estimate $V\simeq 10~\mu\text{eV}$ from the triple point splittings (taking into account the renormalization due to $\tau_C$). We use these values in our NRG calculations, together with the conduction electron bandwidth $D=250~\mu\text{eV}$ (we have verified explicitly that our results are insensitive to further increasing the ratio $D/E_C$, the precise choice of which being somewhat arbitrary). To obtain converged results at larger QPC transmissions, we retained 16 charging states of each island ($\bar{N}=\bar{M}=7$). Comparisons between experiment and theory were carried out at the experimental base temperature of 20 mK.

Although a precise mapping between the QPC transmissions $\tau_{L,R,C}$ and $J_{L,R,C}$ exists in the idealized ballistic limit of noninteracting electrons at a constriction, this approximation was found to be too crude to reproduce even qualitative features of the experiment for this more complex system. Instead, we simply treated the model couplings $J_{L,R,C}$ as free parameters. For a given set of experimental transmissions $\tau_{L,R,C}$ we compared the conductance line cut along the line $U_L=U_R$ to NRG calculations to fix the optimal $J_{L,R,C}$ reported in the main text.

\subsubsection*{Cross-capacitive gate effects}
\begin{figure*}[h]
\centering
\begin{subfigure}{\columnwidth}
  \centering
  \includegraphics[width = 3.2in]{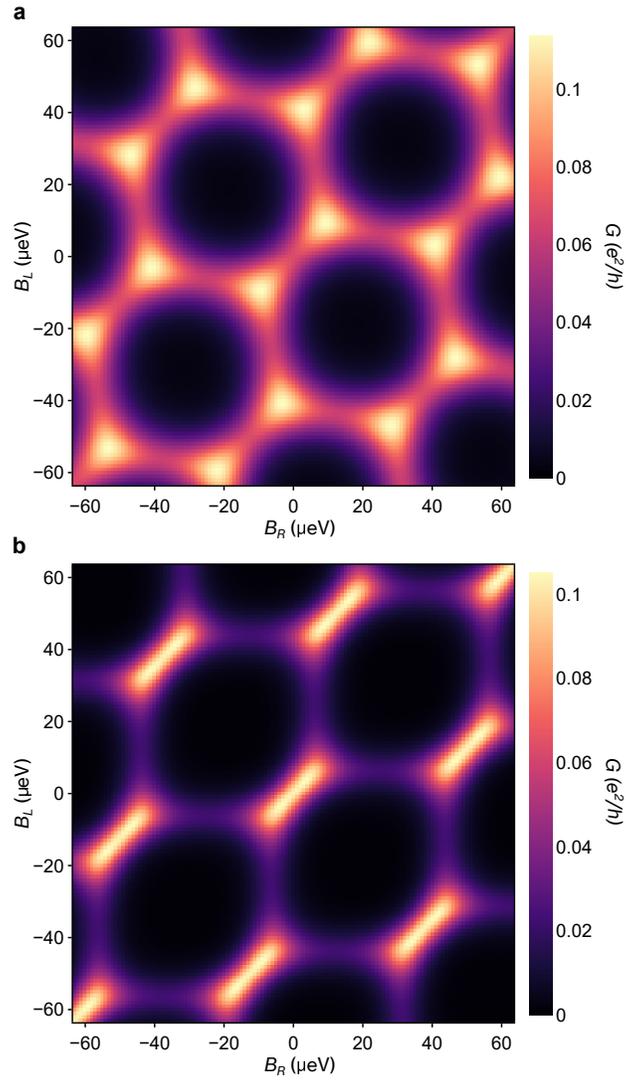}
  \phantomcaption
  \label{fig:expttriple}
\end{subfigure}
\begin{subfigure}{0\columnwidth}
  \centering
    \includegraphics[width=\columnwidth]{RoughDraftFigures/figSuppUnskewedNRG.eps}
    \phantomcaption
    \label{fig:NRGtriple}
\end{subfigure}\ignorespaces
\vspace{-5pt}
\caption{\textbf{Unskewed NRG charge stability diagram.} Charge stability diagrams of Fig.~\ref{fig:stabNRG20} as a function of $B_L, B_R$.}
\label{fig:NRGUnskewed}
\end{figure*}

$B_L$ and $B_R$ in the model can be connected to the experimental parameters $U_L$ and $U_R$ via $\vec{B}=\bar{\boldsymbol{\alpha}}\vec{U}$. Off-diagonal elements of the dimensionless 2x2 matrix $\bar{\boldsymbol{\alpha}}$ correspond to cross-capacitive gate effects. In fitting to the experimental data, we used $\bar{\boldsymbol{\alpha}}_{LL}=\bar{\boldsymbol{\alpha}}_{RR}=1$ and $\bar{\boldsymbol{\alpha}}_{LR}=\bar{\boldsymbol{\alpha}}_{RL}=0.3$. The unskewed conductance colorplot in the space of the NRG parameters $(B_L,B_R)$ corresponding to Fig.~\ref{fig:stabNRG20} is shown for reference in Extended Data Fig.~\ref{fig:NRGUnskewed}.

\subsubsection*{Universal regime at triple point}
In Fig.~\ref{fig:scalingCollapse} we present NRG results in the universal regime near the critical point at the TP. For this, we employ a simpler model involving just the lowest three states of the two islands at this point in the phase diagram, $|N,M\rangle / |N,M+1\rangle / |N+1,M\rangle$. This corresponds to $E_C^{L,R},V \to \infty$.

\subsection*{Relation to two-impurity Kondo model and to ``conventional'' double quantum dots}
In the limit of large charging energies $E_C^{L,R}$, the two-site charge-Kondo device is described by the  DCK model Eq.~\ref{eq:H2dck}, with spin-$\tfrac{1}{2}$ operators for the island charge pseudospins. The DCK model is as such a variant of the celebrated two-impurity Kondo (2IK) model Eq.~\ref{eq:H2ik}, in which two spin-$\tfrac{1}{2}$ quantum impurities are each coupled to their own lead and exchange-coupled together~\cite{Jones1988, Jayaprakash1981, Georges1995, Affleck1992, Simon2005,mitchell2012universal,mitchell2012two}. The 2IK model reads,
\begin{equation}\label{eq:H2ik}
H_{\rm 2IK} = H_{\rm elec} + J_L \vec{\mathcal{S}}_L \cdot \vec{s}_L + J_R \vec{\mathcal{S}}_R \cdot \vec{s}_R + J_C \vec{\mathcal{S}}_L \cdot \vec{\mathcal{S}}_R \;.
\end{equation}
The above 2IK model has some similarities and differences to the DCK model studied in this work, which we elucidate in this section.

On the level of the effective models themselves, the two key differences between Eqs.~\ref{eq:H2dck} and \ref{eq:H2ik} are (i) the exchange coupling terms in the DCK model are anisotropic, containing only the spin-flip terms, while they are SU(2) symmetric in the 2IK model; and (ii) the inter-impurity coupling in the 2IK model is a simple local exchange interaction, while in the DCK model it is a correlated tunneling process involving electrons at the central QPC. In terms of differences in the resulting physics, the coupling anisotropy of the DCK model in (i) is unimportant, since under renormalization \cite{wilson1975renormalization} at low temperatures, the couplings become effectively isotropic, as in the 2IK model. However, the new inter-island coupling term of the DCK model in (ii) makes a dramatic difference. 

Although both DCK and 2IK embody a competition between individual Kondo screening of the spins of the two sites and collective inter-site screening, the latter is a simple two-body \textit{local} singlet in the 2IK model but a many-body \textit{Kondo} singlet spanning both sites in the DCK model. An obvious sign of this difference is that the frustration of screening responsible for quantum criticality sets in at ${J_C \sim T_K}$ for the 2IK model \cite{Affleck1992}, but for the DCK model we have $J_C \sim J_L, J_R$.

At the 2IK critical point, the non-Fermi liquid fixed point properties \cite{Affleck1992} of the 2IK model can be understood in terms of the two-channel Kondo model \cite{mitchell2012two}, in which two relevant degrees of freedom of the two-site cluster are `overscreened' by two spinfull conduction electron channels. By contrast, at the critical triple point of the DCK model we have three collective states of the two-site cluster being overscreened by three effective conduction electron channels. This leads to very different critical properties, including for example a 2IK residual entropy of $S_{\rm imp}=\ln\sqrt{2}$ but $S_{\rm imp}=\ln\sqrt{3}$ in the DCK model, and different conductance signatures (the most stark of which being the $2e^2/h$ maximum conductance with square-root temperature corrections in 2IK, but $e^2/3h$ conductance with $2/3$ power temperature corrections in DCK). To our knowledge, the DCK model supports an entirely new kind of critical point.

The quantum phase transition in the standard 2IK model \cite{Jones1988} is often argued to capture the competition in $f$-electron heavy fermion systems between Kondo screening of local moments by conduction electrons, and magnetic ordering of the local moments driven by a through-lattice RKKY exchange interaction. However, heavy fermion materials contain a lattice of many local moments immersed in a \textit{common} reservoir of mobile electrons, whereas the 2IK considers just two local moments, and two \textit{distinct} electron reservoirs, with a purely local inter-impurity spin exchange. In the 2IK model, the artificial separation into two distinct conduction channels means that its specific critical point is not found in real heavy fermion systems. Furthermore, a similar problem arises when attempting to realize 2IK criticality in ``conventional'' (small, semiconductor) double quantum dot devices, described by an effective two-impurity Anderson model~\cite{jayatilaka2011two}. The very charge fluctuations on the dots required to mediate a series current spoil the independence of the channels, and thereby smooth the 2IK quantum phase transition into a crossover~\cite{Zarand2006}. Indeed, even incipient signatures of 2IK criticality have never been observed experimentally in any double quantum dot or bulk system.

By contrast, the DCK model describing the two-site charge Kondo device does support a quantum phase transition, and distinctive transport signatures of it have been observed in our experimental setup. Furthermore, we argue that the \textit{collective} many-body screening of the two sites mediated by conduction electrons in our system is more directly analogous to the low-temperature development of lattice coherence in real bulk heavy fermion materials. 

Finally, we comment on the differences between our two-site charge-Kondo device and standard double quantum dots on scanning plunger gate voltages over the full phase diagram. In our device, we see a regular hexagonal charge stability diagram (Fig.~\ref{fig:device}), with relatively high conductance along lines separating different charge configurations on the two sites (with the highest conductance being at the triple point). As the temperature is decreased, the inter-site Kondo effect works to \textit{suppress} the series current through the device between the triple points, while the conductance at the triple points is relatively enhanced by Kondo. By contrast in the standard double quantum dot, there is no Kondo renormalization on lowering the temperature between the triple points. Furthermore, there is a clear odd-even filling effect in standard spinfull double quantum dots, with Kondo-boosted conductance only manifest when one or other of the dots has an odd number of electrons and carries a net spin-$\tfrac{1}{2}$. Since the physical electrons are actually effectively spinless in the DCK system, and Kondo physics arises due to charge degeneracies rather than spin degeneracies, there are no odd-even effects and every hexagon in the charge stability diagram is equivalent to every other one.


\subsection*{Outlook for scaling to more complex clusters}
Technical improvement of the materials platform for devices could be crucial to enable scaling to larger clusters. 
Switching to InAs-based quantum wells would allow boosting the charging energy~\cite{Albrecht2016} -- and thus Kondo temperature -- by shrinking the islands, without sacrificing the other demanding requisites such as well-defined quantum Hall edge states, high-transparency of miniature ohmic contacts to those edges, and quantum point contacts that smoothly tune transmission of the edge modes. This would increase our window of temperature over which conduction could be described as universal, and improve our ability to isolate the behavior near critical points even at strong coupling. As more islands are coupled together, measuring near a particular set of near-degenerate charge states, with limited influence from higher-energy states, becomes even more important and challenging.

\clearpage

\begin{figure*}[h]
\centering
\includegraphics[width = 3.2in]{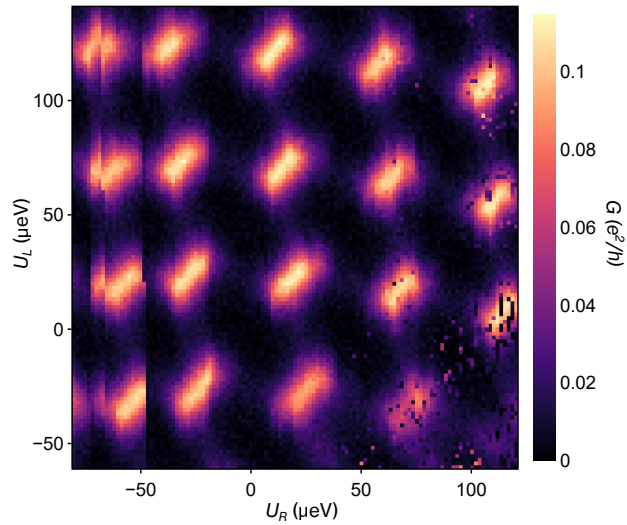}
\caption{\textbf{Periodicity of charge stability diagram.} The bottom charge stability diagram of Fig.~\ref{fig:stabExpt} over a broader range of $U_L, U_R$ showing the periodicity. This also shows the two representative types of charge noise which affect measurements -- occasional jumps (on the left side) where there is a clear vertical discontinuity, and more scattered individual points of noise (on the far right).
}
\label{fig:HoneycombFull}
\end{figure*}
\clearpage

\end{document}